%% file: main.tex
\documentclass[fleqn,usenatbib]{mnras}

\usepackage{newtxtext,newtxmath}

\usepackage[T1]{fontenc}

\DeclareRobustCommand{\VAN}[3]{#2}
\let\VANthebibliography\thebibliography
\def\thebibliography{\DeclareRobustCommand{\VAN}[3]{##3}\VANthebibliography}

\usepackage{graphicx}	
\usepackage{amsmath}	
\usepackage{threeparttable}  
\usepackage{booktabs} 

\title[Variable stars in Eridanus II]{Variable stars in Local Group Galaxies - V. The fast and early evolution of the low-mass Eridanus II dSph galaxy}

\author[C. E. Mart\'{i}nez-V\'{a}zquez et al.]{
C.~E. Mart\'{i}nez-V\'{a}zquez,$^{1}$\thanks{E-mail: clara.martinez@noirlab.edu (CEMV)}
M. Monelli,$^{2,3}$
S. Cassisi,$^{4,5}$
S. Taibi,$^{2,3}$
C. Gallart,$^{2,3}$
A.~K. Vivas,$^{1}$
\newauthor
A.~R. Walker,$^{1}$
P. Mart\'{i}n-Ravelo,$^{6}$
A. Zenteno,$^{1}$ 
G. Battaglia,$^{2,3}$
G. Bono,$^{7,8}$
A. Calamida,$^{9}$
\newauthor
D. Carollo,$^{10}$
L. Cicu\'{e}ndez,$^{2,3}$
G. Fiorentino,$^{8}$,
M. Marconi,$^{11}$
S. Salvadori,$^{12, 13}$
\newauthor
E. Balbinot,$^{14}$
E.~J. Bernard,$^{15}$
M. Dall'Ora,$^{11}$
P.~B. Stetson$^{16}$\\
\\
$^{1}$ Cerro Tololo Inter-American Observatory, NSF's NOIRLab, Casilla 603, La Serena, Chile\\
$^{2}$ IAC-Instituto de Astrof\'isica de Canarias, Calle V\'ia Lactea s/n, E-38205 La Laguna, Tenerife, Spain\\
$^{3}$ Departmento de Astrof\'isica, Universidad de La Laguna, E-38206 La Laguna, Tenerife, Spain\\
$^{4}$ INAF-Osservatorio Astronomico d'Abruzzo, via M. Maggini, sn. I-64100 Teramo, Italy\\
$^{5}$ INFN-Sezione di Pisa, Largo Pontecorvo 3, I-56127 Pisa, Italy\\
$^{6}$ Departamento de Astronom\'ia, Universidad de La Serena, Av. Juan Cisternas 1200 Norte, La Serena, Chile\\
$^{7}$ Department of Physics, Universit\`a di Roma Tor Vergata, via della Ricerca Scientifica 1, 00133 Roma, Italy\\
$^{8}$ INAF-Osservatorio Astronomico di Roma, via Frascati 33, 00040 Monte Porzio Catone, Italy\\
$^{9}$ Space Telescope Science Institute, 3700 San Martin Drive, Baltimore, MD 21218\\
$^{10}$ INAF-Osservatorio Astronomico di Trieste, via G. B. Tiepolo 11, I-34143 Trieste, Italy\\
$^{11}$ INAF-Osservatorio Astronomico di Capodimonte, Via Moiariello 16, 80131 Napoli, Italy\\
$^{12}$ Dipartimento di Fisica e Astrofisica, Univerisit\`a degli Studi di Firenze, via G. Sansone 1, Sesto Fiorentino, Italy\\
$^{13}$ INAF-Osservatorio Astrofisico di Arcetri, Largo E. Fermi 5, Firenze, Italy\\
$^{14}$ Kapteyn Astronomical Institute, University of Groningen, Postbus 800, NL-9700AV Groningen, The Netherlands\\
$^{15}$ Universit\'e C\^ote d'Azur, OCA, CNRS, Lagrange, France\\
$^{16}$ Herzberg Astronomy and Astrophysics, National Research Council Canada, 5071 West Saanich Road, Victoria, BC V9E 2E7, Canada \\
}

\date{Accepted 2021 August 26. Received 2021 August 25; in original form 2021 February 24}

\pubyear{2021}

\begin{document}
\label{firstpage}
\pagerange{\pageref{firstpage}--\pageref{lastpage}}
\maketitle

\begin{abstract}

We present a detailed study of the variable star population of Eridanus~II (Eri~II), an ultra-faint dwarf galaxy that lies close to the Milky Way virial radius. We analyze multi-epoch $g,r,i$ ground-based data from Goodman and the Dark Energy Camera, plus $F475W, F606W, F814W$ space data from the Advanced Camera for Surveys. We report the detection of 67 RR Lyrae (RRL) stars and 2 Anomalous Cepheids, most of them new discoveries. With the RRL stars, we measure the distance modulus of Eri~II, $\mu_0=22.84\pm 0.05$~mag (D$_{\odot}=370\pm9$~kpc) and derive a metallicity spread of 0.3~dex (0.2~dex intrinsic). The colour distribution of the horizontal branch (HB) and the period distribution of the RRL stars can be nicely reproduced by a combination of two stellar models of [Fe/H]=($-2.62$, $-2.14$). The overall low metallicity is consistent with the red giant branch bump location, 0.65~mag brighter than the HB. These results are in agreement with previous spectroscopic studies. The more metal-rich RRL and the RRab stars have greater central concentration than the more metal-poor RRL and the RRc stars that are mainly located outside $\sim 1$~r$_{\rm h}$. This is similar to what is found in larger dwarf galaxies such as Sculptor, and in agreement with an outside-in galaxy formation scenario. This is remarkable in such a faint dwarf galaxy with an apparently single and extremely short ($<1$~Gyr) star formation burst. Finally, we have derived new and independent structural parameters for Eri~II and its star cluster using our new data that are in very good agreement with previous estimates.

\end{abstract}




\section{Introduction}
Old stellar populations ($> 10$ Gyr) provide fundamental constraints on the conditions of the early Universe and the physics governing the early stages of galaxy evolution \citep[e.g.,][]{Bernard2008, Stetson2014, MartinezVazquez2016a, Fiorentino2017, Vivas2020a}. Although we still lack accurate measurements of the dynamical mass of RR Lyrae (RRL) stars, current circumstantial evidence suggests that they are low mass stars ($\sim 0.7M_{\odot}$) in their core He-burning horizontal branch (HB), implying ages larger than $\approx 10$~Gyr \citep{Walker1989, Savino2020}. This has been thoroughly confirmed empirically, though rare classes of RRL stars have been identified when other physical mechanisms are at play, such as binarity \citep{Pietrzynski2012}. 

RRL stars are primary distance indicators \citep[e.g., ][]{Beaton2018}, but can also be used to trace the early chemical evolution of the host system \citep{Jeffery2011, Braga2016, MartinezVazquez2016a} because their pulsational parameters are linked to their metal content \citep[e.g.,][]{Jurcsik1996, Alcock2000, Nemec2013, Marconi2015}. The intrinsic brightness ($M_V \approx 0.6$~mag) and the large pulsational amplitude ($\langle \Delta V \rangle= 0.8$~mag) make it possible to study RRL stars to beyond the edge of the Local Group (LG) using Hubble Space Telescope (HST) data \citep[e.g.,][]{DaCosta2010,Yang2014}. Therefore, the properties of RRL stars help to constrain the early properties of galaxies more distant than the limit of current spectroscopic facilities. Moreover, comparative studies of RRL stars in different galaxies provide a unique opportunity to investigate the conditions of the early LG, opening the possibility of a global study that probes beyond the details of individual galaxies \citep{Fiorentino2015a, Fiorentino2017, Monelli2017, MartinezVazquez2017}.

Our view of the LG has drastically changed in the last two decades. In particular, the number of satellites of both the Milky Way (MW) and M31 has increased by almost an order of magnitude thanks to new deep, wide photometric surveys \citep[e.g.,][]{mcconnachie09,DESCollaboration,Drlica-Wagner2016,Drlica-Wagner2021}. While the majority of new stellar systems extend the range of galactic properties to low mass, low luminosity, and low mean metallicity \citep[e.g.,][]{Simon2019}, a few extended galaxies of modest luminosity but with remarkably low surface brightness have been discovered, in particular Crater~II, Antlia~II \citep{Torrealba2016a, Torrealba2018}, And~XXXI, and And~XXIII \citep{Martin2013a}. RRL stars have been found in almost all the galaxies where they have been searched for, with the exception of only some of the lowest-mass ultra-faint dwarf galaxies \citep[Willman1 and Carina III, see the compilation made in][Table~A1]{MartinezVazquez2019}.

In this work we focus on the Eridanus~II (Eri~II) ultra-faint dwarf, discovered from a search of the first year data release of the Dark Energy Survey (DES) \citep{Bechtol2015,Koposov2015}. Eri~II is the presently most distant known possible Milky Way (MW) satellite, at D$\sim$370~kpc \citep{Crnojevic2016b}. It is approaching the MW \citep{Li2017} and possibly bound \citep{Fritz2018}, though the error bars on the Gaia Data Release 2 (DR2) proper motion are too large to set firm constraints.

The spectroscopic investigation of bright red giant branch (RGB) stars \citep{Li2017} supports that Eri~II is a metal-poor system ($\langle$[Fe/H] $\rangle \sim -2.4$) which presents a non-negligible metallicity dispersion ($\sigma_{\rm{[Fe/H]}} = 0.47$). Contrary to other distant satellites such as Leo~I, Leo~II, and Leo~T, Eri~II shows no evidence of an intermediate-age population. This was confirmed by the deep Advanced Camera for Surveys (ACS) photometry and the detailed star formation history analyses recently published \citep{Simon2021,Gallart2021}. Indeed, Eri~II seems to have formed the vast majority of its stars early (before $z\sim$6), with a possible residual population not younger than 9~Gyr.

A remarkable feature of Eri~II is the presence of an apparent stellar overdensity adjacent to its centre \citep{Koposov2015}. This was interpreted as a stellar cluster by \citet{Crnojevic2016b}, making Eri~II the faintest galaxy ($M_V=-7.1$) known to host a cluster, a conclusion supported by the recent photometric \citep{Simon2021} and spectroscopic \citep{Zoutendijk2020} studies. Due to the small number of stars associated with the cluster, no statistically significant age difference can be proven with respect to the main field population.

In this paper, we analyze both ground-based and space data to study the variable star population of Eri~II and its early chemical enrichment based on its RRL star population. We report the discovery of 67 RRL stars and 2 Anomalous Cepheids (AC). The paper is organized as follows. In \S~\ref{sec:data} we present the observing material, data reduction and photometric calibration. In \S~\ref{sec:populations} we derive structural parameters for Eri~II and its tentatively cluster (\S~\ref{sec:structure}), and we discuss some features about the colour-magnitude diagram (CMD, \S~\ref{sec:cmd}). In \S~\ref{sec:vars}, we present the detection, classification, and characterization of variable stars. AC and RRL stars are discussed in detailed in \S \ref{sec:ac} and \S~\ref{sec:rrl}, respectively. In particular, we derive the distance to Eri~II and its metallicity distribution using the population of RRL stars. In \S~\ref{sec:simul} we perform a simulation of the HB to support the spread in metallicity observed in this galaxy. In \S~\ref{sec:discussion}, we discuss the fast and early evolution of Eri~II based on a careful analysis of its RRL stars. Finally, the summary and final remarks of this work (\S~\ref{sec:conclusions}) close the paper. 

\section{Observations and data reduction}\label{sec:data}

The present work uses ground-based and space data. The details of each particular data set are given in the following subsections.

\begin{table}
\small
\caption{Number of epochs collected in Eri~II}
\label{tab:epochs}
\begin{tabular}{lcccccc}
\toprule
Data Source &  N$_g$ & N$_r$ & N$_i$ & N$_{F475W}$ & N$_{F606W}$ & N$_{F814W}$ \\
\midrule
ACS     & & & & 8 & 10 & 24 \\ 
Goodman & 31* & 30* & 30* & & & \\
DECam   &  -- &  4 & 16 & & & \\
\bottomrule
\end{tabular}
\begin{tablenotes}
\item \textit{Notes.-} *This is the number of epochs collected for Eri~II independently in the two fields of Eri~II (East \& West) observed with Goodman.
\end{tablenotes}
\end{table}

\subsection{Goodman and DECam data}\label{sec:ground}

\begin{figure*}
    \includegraphics[width=0.7\textwidth]{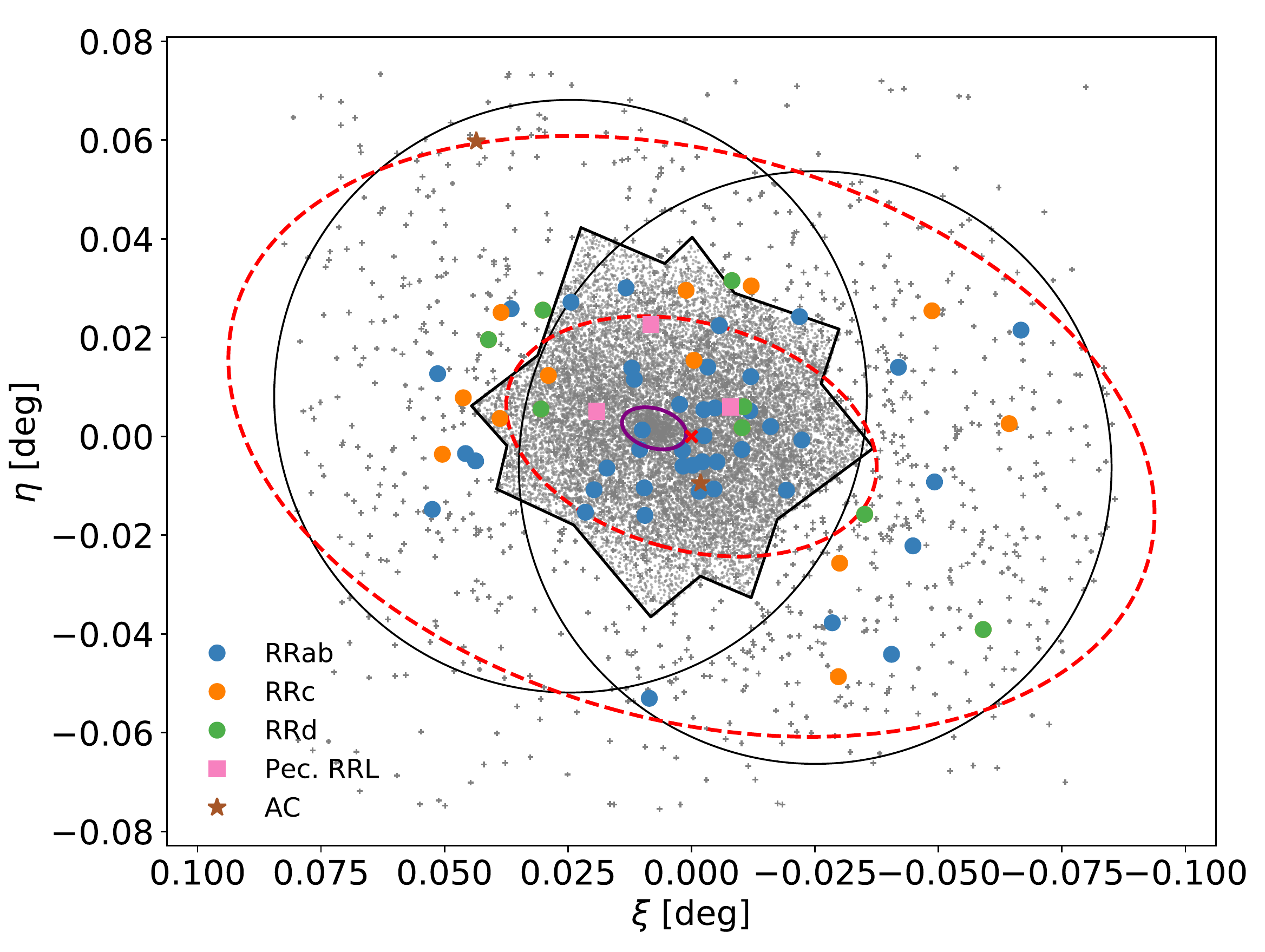}
    \caption{Spatial distribution (in planar coordinates) of the variable stars detected in Eri~II. The Goodman+DECam photometry of non-variable stars is shown by grey crosses while the ACS photometry of non-variable stars is identified with grey dots. The field-of-view of the ACS data for Eri~II is outlined by black lines and the locus of the Goodman pointings are represented by black circles. The 1~r$_{\rm h}$ locus of the central star cluster in Eri~II is represented by the purple ellipse. The dashed red ellipses mark the locus corresponding to 1~r$_{\rm h}$ and 2.5~r$_{\rm h}$ of Eri~II using the morphological parameters calculated in this work.}
    \label{fig:spatial}
\end{figure*}

Ground-based data were taken under NOAO PropID. 2018A-0310 (PI. C.~E. Mart\'inez-V\'azquez). We collected multi-epoch data using the Goodman High Throughput Spectrograph \citep{Clemens2004} (red camera), in imaging mode, at the 4m Southern Astrophysical Research (SOAR) Telescope, located on Cerro Pach\'on (Chile). The Goodman imager is characterized by a circular field of view (FoV) of 7.2\arcmin diameter with a 0.15\arcsec/pixel scale. In order to reduce readout time and increase the signal to noise, we selected $2 \times 2$ binning in our observations. Our strategy consisted in taking two overlapping Goodman pointings in the East (RA = 03:44:28.0, Dec = $-$43:31:36.7) and West (RA = 03:44:11.6, Dec = $-$43:32:28.7) side of Eri~II, covering the body of Eri~II to more than 2~r$_{\rm h}$, and doubling the number of measurements in the innermost regions. We obtained time series photometry in $g$, $r$, and $i$. The individual exposure times were between 120 and 600~s. The cadence of these observations was optimized for RRL stars, taking into account the actual weather conditions, seeing, and the moon phase and position. The images were acquired during non-consecutive nights: October 31st, November 1st and November 30th, 2018. This cadence helps to minimize the aliasing in the period determination of RRL stars with $P \approx 0.5$~d. In order to prevent having a saturated background due to the bright Moon and to allow easier removal of cosmic ray contamination, we split long exposures into 2-3 shorter exposures, in the same filter. Due to the short readout time of the camera (10.3 seconds) this procedure did not introduce significant inefficiency. The total number of epochs gathered in $g$, $r$, and $i$ for each pointing are given in Table~\ref{tab:epochs}. The images were collected under grey conditions (Moon phase $\sim 6$~d) with a median value of the seeing of 0.77\arcsec.

During each run, a set of bias exposures and dome flats were collected in the afternoon. In addition, sky flats were secured during the evening twilight (only for the Goodman data set; sky flats at DECam are forbidden). The set of biases and flats that were closest in time were used for processing each exposure. Images were corrected using {\sc iraf}\footnote{{\sc iraf} \citep{Iraf1, Iraf2} was distributed by the National Optical Astronomy Observatory, which was managed by the Association of Universities for Research in Astronomy (AURA) under a cooperative agreement with the National Science Foundation.} tasks for bias subtraction and flat-fielding. We followed exactly the same procedure detailed in \citet{MartinezVazquez2019} (their Section 3.1). 

Additional data were taken in bright time using DECam \citep{Flaugher2015}, a wide FoV camera (3 deg$^2$, 62 science CCDs, 0.263\arcsec/pixel) installed at the prime focus of the V\'ictor Blanco 4-m telescope at Cerro Tololo Inter-American Observatory (CTIO) in Chile. The data were taken during small time windows available during Director's Discretionary Time, on average one image every few days for a total time baseline of $\sim$90 days between mid November 2018 and mid February 2019. The galaxy was centered on DECam chip S4 at the center of the field of view. The entire galaxy (r$_{\rm h}$ = 2.3\arcmin, see Table~\ref{tab:morph_param}) fits well within a single CCD, which has a FoV of 9\arcmin $\times$ 18\arcmin. Because these nights were close to Full Moon, we mostly collected $i$ band data, since this band is less affected by the increase of the sky background due to the brightness of the Moon. 

The basic reduction was performed using the DECam Community Pipeline \citep{Valdes2014} for bias, flat-fielding, illumination correction, and astrometry. These data were taken in groups of four consecutive images, with individual exposure times of typically 250~s to avoid background saturation. In order to increase the S/N for the Eri~II RRL stars, we summed all the good quality images of each block to produce one phase point in each filter, thus resulting in a total of 16 and 4 individual epochs in $i$ and $r$ filters, respectively.

Figure~\ref{fig:spatial} shows the distribution of sources detected in the observed field. Eri~II was well covered by just one of the DECam CCDs. The distribution of points corresponds to a portion of CCD S4 in DECam, while the two SOAR/Goodman pointings are shown by the two red circles (see \S \ref{sec:ground}). The plot also shows the location of the detected variable stars (\S~\ref{sec:vars}).

\begin{figure}
	\centering
	\includegraphics[width=0.4\textwidth]{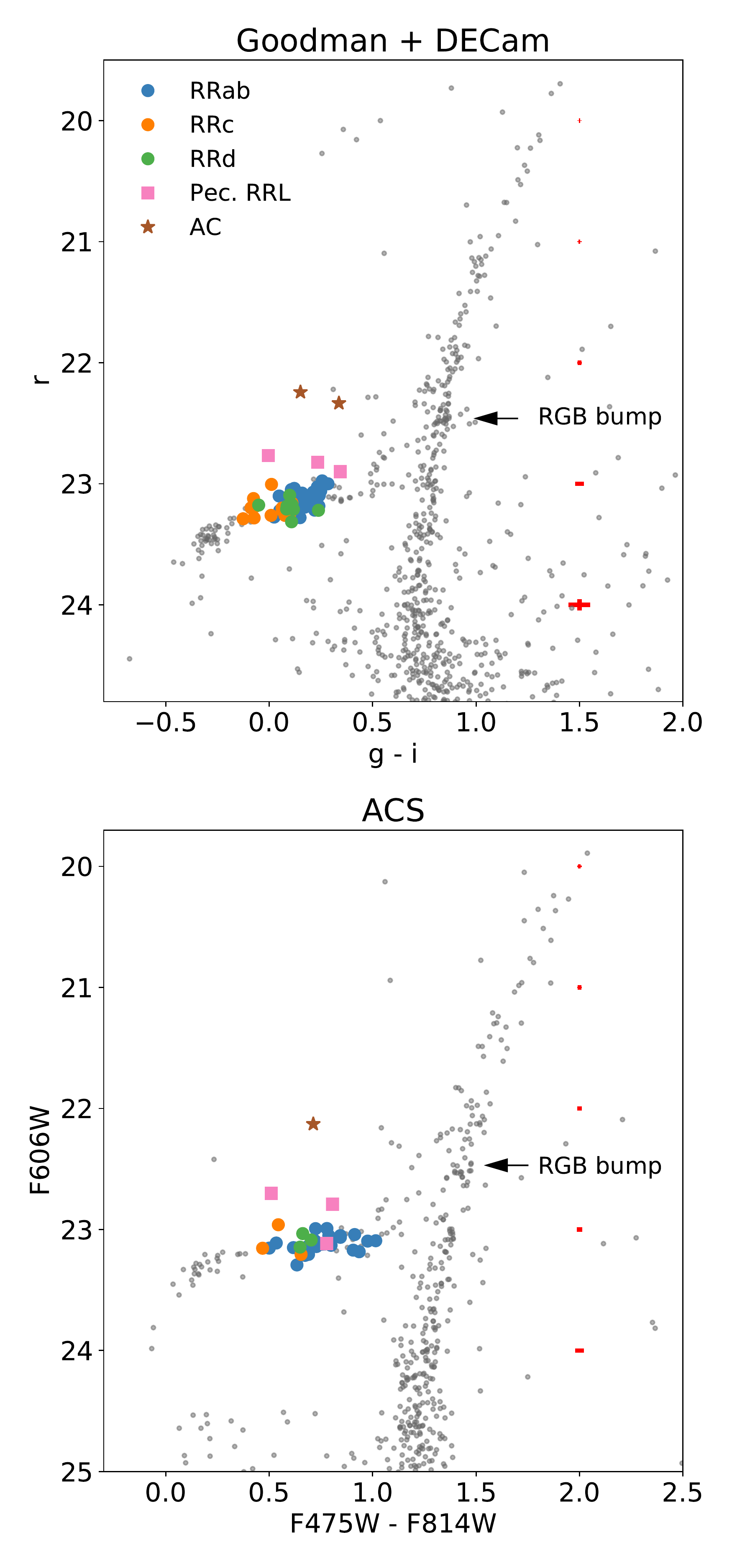}
	\caption{CMDs of Eri~II using Goodman+DECam (top panel) and ACS (bottom panel) data. Variable stars detected in this work are shown with different colours and symbols (same as in Figure~\ref{fig:spatial}). The arrows point the location of the RGB bump.}
	\label{fig:cmd}
\end{figure}

\subsection{ACS data}\label{sec:acs}

HST images have been collected by two independent projects in cycle 23, GO 14234 (P.I. Simon, 13 orbits) and GO 14224 (P.I. C. Gallart, 6 orbits), see also \citet{Gallart2021} for further details. The first program was executed between January 6th and February 2nd, 2016, thus covering a total of 20.8~d. A total of 10 and 16 images in the $F606W$ and $F814W$ filters were collected, with exposure times ranging between 1,220 and 1,390~s.  The observing strategy was such that images in one filter were observed during each orbit, and the bulk of observations in each filter were grouped together. Therefore, the data cover a time range of 14.3~d and 7.6~d, in the $F606W$ and in the $F814W$ bands, respectively. The second program, observed on September 22nd and 23rd, 2016, had a different strategy. In order to optimize the search for short period variable stars such as RRL stars and ACs, each orbit was split into one $F475W$ and one $F814W$ image. Six images of 1,257s and 1,300~s, respectively, plus two short exposures were collected. The observations consisted of two blocks of $\sim$3.5~hr, separated by a gap of 24~hr, thus covering in total 1.3~d. The two data sets, collected in February and September 2016, were therefore separated by $\approx$236~d. Moreover, a rotation of about $\simeq 121$ degrees is present between the two sets. This means that sources in the innermost regions have been observed with the three filters, while the region close to the corners of the camera have been covered by two filters only, with only $F814W$ covering the whole HST field of view. Table~\ref{tab:epochs} shows the total number of epochs observed from both data sets in the $F475W$, $F606W$, and $F814W$ filters.

Figure~\ref{fig:spatial} shows the area covered by the two HST programs. The field-of-view of the two ACS pointings is enclosed by black lines, where it is clearly visible how the ACS photometry is significantly deeper than that from the ground-based data.

\subsection{Instrumental photometry and photometric calibration}\label{sec:reduction}

The data reduction was performed using the {\sc DAOPHOT-IV/ALLFRAME} programs \citep{Stetson1987, Stetson1994}, following the prescriptions by \citet{Monelli2010b}. The two ACS and the two ground-based data sets were reduced separately, and combined \textit{a posteriori}. In the case of the DECam images we only processed the CCD S4 since it covers completely the area of the two Goodman pointings and thus the expected extension of Eri~II (as shown in Figure~\ref{fig:spatial}).

We then adopted the following procedure: As a first step, an iterative run of the {\sc DAOPHOT} and {\sc ALLSTAR} routines are used to identify the best list of PSF stars. Subsequently, {\sc ALLFRAME} processes all images simultaneously, using a list of stars detected in a registered, combined median image. The full procedure is iterated twice to exploit the improved knowledge of the stellar centroid and coordinate transformations to refine both the PSF on individual images and the star list. The process provides a catalogue of sources with instrumental magnitude for each available image.

Additionally, we mention that the parallel WFC3 fields were reduced as well with an identical approach, but since no variable stars were identified in either data set, they will not be discussed further.

\subsection{Photometric calibration}\label{sec:calib}

To build the time series data, we calibrated the individual ground-based catalogues to the standard DES photometric system, as defined by the DES DR2 catalogue\footnote{\url{https://des.ncsa.illinois.edu/releases/dr2}}, which has a photometric precision of the order of 2~mmag in the $g,r,i$ bands and a median depth of $g = 24.7$, $r = 24.4$, $i = 23.8$~mag at $S/N = 10$ \citep{DES_DR2}. The transformation equations between the instrumental and the $g,r,i$ DES magnitudes were derived only for those stars with magnitude uncertainties less than 0.05 mag, obtaining zero-points, colour terms and linear positional terms (X and Y). The positional terms were negligible and the colour term was only applied to the Goodman observations. The DECam instrumental magnitudes do not show any colour dependency when comparing with DES, which is expected since DES uses the same DECam passbands. The root-mean-square deviations of the transformations from the instrumental to the calibrated magnitudes were 0.03~mag in $r,i$-DECam, 0.05~mag in $g$-Goodman and 0.04~mag in $r,i$-Goodman. The $g,r,i$-Goodman filters are very close to the $g,r,i$-SDSS passbands. Our catalogue depth at S/N = 10 is $g,r,i$ = (25.2, 24.5, 24.2)~mag.

In the case of the ACS data, we started from the photometry catalogues for individual images obtained from {\sc ALLFRAME} \citep{Gallart2021},  individually calibrated to the VEGAMAG system (see Figure \ref{fig:cmd}, bottom panel).

\section{Properties of the stellar populations of Eri~II}\label{sec:populations}

\subsection{Structural parameters of Eri~II and its cluster}\label{sec:structure}

\input{morph_param_tab.tex}

To derive the global structural parameters of Eri~II, we fitted an exponential function plus an uniform distribution of contaminants, to our Goodman+DECam data. Although these data already resulted in a relatively clean CMD (see Figure~\ref{fig:cmd}, top panel), we performed an additional selection retaining those stars within $19<i<24$ (i.e., just below the tip of the RGB and the HB, respectively) and $(g-i)<1.5$, so to remove residual foreground and background contamination sources. 

To derive the structural parameters, we followed the same methodology by \citet{Cicuendez2018a} based on \citet{Richardson2011}. The procedure evaluates, at the individual stellar locations, a likelihood expression depending on the surface density profile to be fitted as well as on the position angle, ellipticity, and the density of contaminants. The most likely value of each structural parameter is then evaluated by sampling this likelihood expression with a Markov chain Monte Carlo (MCMC) method. We used a Bayesian MCMC sampler with affine invariance of the form given by \cite{Goodman2010}: `The MCMC Hammer' \citep{Foreman-Mackey2013}\footnote{The code used was \url{https://github.com/grinsted/gwmcmc}}. We established a total of 80 walkers, each of them performing 6000 steps.

Table~\ref{tab:morph_param} shows the structural parameters obtained for Eri~II.
Our results are in very good agreement with those reported by \citet{Crnojevic2016b}, confirming the goodness of their finding. In the same way, we also agree with \citet{Simon2021}, although our half-light radius and the right-ascension of the galaxy's central coordinates are significantly different (at 3-$\sigma$ level) than their reported values. It is probable that the half-light radius reported by \citet{Simon2021} was overestimated due to the small field-of-view of their HST data, as these authors explained when comparing their results with those of \citet{Crnojevic2016b}. We therefore attribute to the same reason the difference we found in the central coordinates of Eri~II.

Since the discovery by \citet{Crnojevic2016b}, the existence of an off-center stellar cluster in Eri~II has attracted a lot of attention, due to the low-mass of Eri~II and the possibility that the cluster would help characterize its dark matter content \citep{Amorisco2017,Contenta2018,Marsh2019,Simon2021}. We therefore used our 3-bands HST dataset, which covers a smaller portion of the galaxy but with deeper photometry, to fit the surface density profile of the candidate star cluster by modeling its background contamination density with the constrained fitted profile of Eri~II plus a uniform distribution to account for contamination by foreground MW stars. Table~\ref{tab:morph_param} shows the structural parameters obtained for the cluster, comparing with the results reported in \citet{Crnojevic2016b} and \citet{Simon2021}. 

Data listed in this table show that Eri~II and its candidate star cluster share the same ellipticity and position angle, within their respective error bars, a surprising finding already noted by \citet{Simon2021}. The center of the cluster presents a small but significant offset with respect to the galaxy center, with the cluster placed $32\arcsec\pm5\arcsec$ to the East to the galaxy centre, which translate to $57\pm 9$~pc, assuming our calculated heliocentric distance reported in \S~\ref{sec:distance}. This value is comparable to what found by \citet{Crnojevic2016b} (i.e., $\sim25\arcsec$), but significantly larger (at 3-$\sigma$ level) than the value reported by \citet{Simon2021} (i.e., $13.9\arcsec\pm2.0\arcsec$ or $23\pm3$~pc). Since the cluster central coordinates that we found are almost identical to those of \citet{Simon2021} (who used similar HST data), we attribute this discrepancy to the different central coordinates for Eri~II. 

We note that a larger offset between the cluster and the galaxy center would imply a larger core radius for the dark matter halo than previously reported by \citet{Simon2021}. Qualitatively, if we assume, as these authors did (see their Section~5.3 for details), that the calculated offset is $\sim 1/3$ of the core radius, then this would result in a core radius $\sim 170$~pc. Therefore, the core radius of Eri~II would remain small, but now closer to those values postulated for other dwarfs, i.e., ranging from a few hundred pc to one kpc.
 
Despite the very low stellar density, the kinematic analysis by \citet{Zoutendijk2020} supports that the cluster is real, as the velocity dispersion of its stars is smaller than that of stars in the main body of Eri~II. On the other hand,  \citet{Simon2021} tried to estimate the cluster age, but given the very low number of stars in the cluster regions, they could only find marginal evidence that the cluster is slightly younger ($\sim$0.6~Gyr) than the bulk of the Eri~II population.

\begin{figure}
    \centering
    \includegraphics[width=0.4\textwidth]{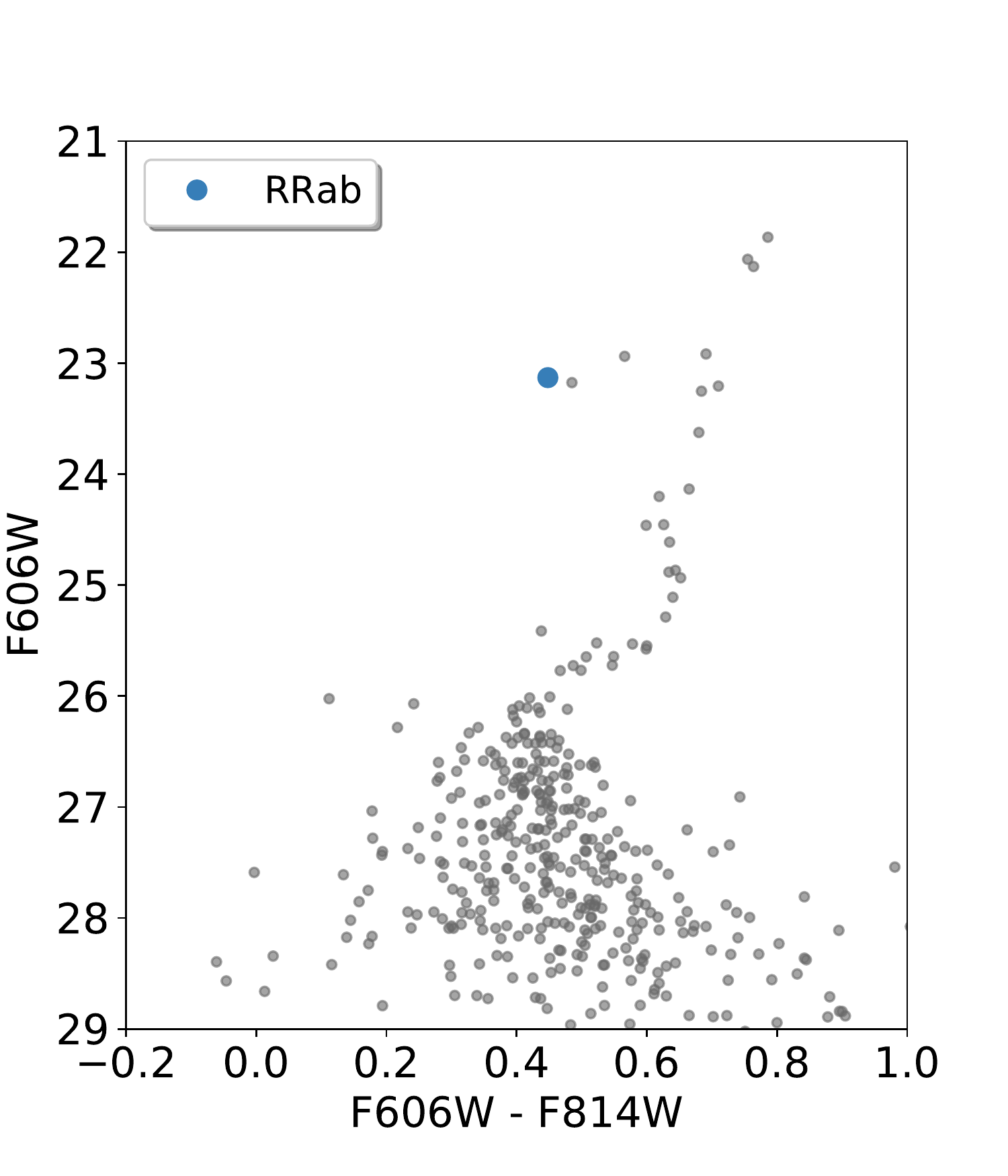}
    \caption{CMD of stars within 1\,r$_h$ Eri~II cluster.}
    \label{fig:cmd_cluster}
\end{figure}

To check whether the analysis of the variable stars can add information about this, Figure \ref{fig:spatial} shows the location of the detected variable stars, shown with filled symbols as labelled, as well as the estimated center of Eri~II (red cross) and the cluster's half-light radius (purple ellipse). The comparison discloses that the center of the distribution of the RRL stars is close to the Eri~II center, separated by only $\sim$6\arcsec as reported in Table \ref{tab:morph_param}. We find that there is no accumulation of RRL stars associated with the cluster, and only one RRL star (V46, see \S~\ref{sec:vars}) is located with 1~r$_{\rm h}$ from the cluster's center. The analysis of the CMD of the stars within the 1~r$_{\rm h}$ from the cluster's center reveals that only two stars appear to be on the HB, the aforementioned RRL star plus one red HB star (see Figure~\ref{fig:cmd_cluster}).

\subsection{The Colour-Magnitude Diagram and the Red Giant Branch}\label{sec:cmd}

Figure \ref{fig:cmd} shows the ground-based $r, (g -i)$ (top panel) and the ACS $F606W, (F475W-F814W) $ (bottom panel) CMDs, highlighting the detected variable stars. Notably, the Goodman+DECam CMD reaches $\sim$1.5~mag fainter that the HB, to $r \sim 25.0$~mag. The CMD was cleaned by visual inspection of the images, removing background galaxies. A detailed discussion of the main features of the Eri~II ACS CMD has been presented in \citet{Gallart2021}, and will not be repeated here. Nevertheless, two features are worth discussing. First, we confirm the colour range of HB stars, on both side of the instability strip. Interestingly, the blue part is significantly more populated in the ground-based CMD, which covers a wider area. A detailed discussion of the spatial variation of the HB morphology is presented in \S \ref{sec:discussion}. Secondly, we report here the detection of the RGB bump, which was not analysed in previous works. This appears as an over-density in the luminosity function of the RGB, caused by the H-burning shell crossing the chemical discontinuity left behind the convective envelope (see e.g., \citealt{cs:97} and references therein). This results in a re-adjustment of the stellar  structure that leads to a decrease, and then an increase in luminosity. This is reflected by an overdensity in the star counts up the RGB, due to the stars crossing the same region of the CMD three times. The position of the RGB-bump depends both on the age and the metallicity, becoming fainter as either of these two parameters increase. Therefore, the RGB bump position is a useful observable to investigate the envelope chemical stratification of low-mass stars, and the occurrence of non-canonical mixing processes at the bottom of the convective envelope \citep[see, e.g.,][and references therein]{csb:02}. The magnitude difference between the RGB bump and the HB luminosity, being distance and reddening independent, has been extensively adopted to test the reliability of the stellar evolutionary framework. The latest investigations \citep{Dicecco2010a,Monelli2010c,Cassisi2011} indicate an overall agreement between theory and observations, but have consistently shown a disagreement in the most metal-poor regime ([M/H]$<$--1.5), in the sense that the observed location of the bump is fainter than the predicted one.

From analyses of the different CMDs at our disposal, we have consistently detected the RGB bump in the different filters. In particular, the peaks of the RGB luminosity function, corresponding to the RGB bump, were detected at $g_{\rm bump}=23.11\pm0.05$~mag, $r_{\rm bump}=22.46\pm0.05$~mag, $i_{\rm bump}=22.24\pm0.05$~mag, and $F606W_{\rm bump}=22.47\pm0.05$~mag. The location of the bump in $r$ and $F606W$ is indicated by the arrows in Figure \ref{fig:cmd}. Since the bump gets brighter with decreasing metallicity, its location (well above the magnitude level of the HB) is an independent qualitative indicator that the mean metallicity is low, in agreement with spectroscopic measurements \citep{Li2017}. 
To further investigate the bump properties, we take advantage of the ACS data and follow the approach proposed by \citet{Monelli2010c}. This is based on defining a pseudo-$V$ magnitude as $V^{\ast}$=($F475W+F814W$)/2, which provides HB morphology which is horizontal in the colour range of RRL stars, facilitating the estimate of its magnitude level. In particular, the luminosity function of the RRL stars is fitted with a Gaussian distribution, and the lower envelope level, representative of the ZAHB locus, is calculated to be 1.5$\sigma$ fainter than the peak of the magnitude distribution. This analysis gives $V^{\ast}_{\rm HB}$=23.21$\pm$0.03~mag which, combined with the bump location at $V^{\ast}_{\rm bump}$=22.56$\pm$0.03 provides $\Delta V_{\rm HB}^{\rm \ast bump}\,= -0.65$~mag. 
At face value, such estimate would imply either an age much younger than 10~Gyr (for [Fe/H]=$-2$) or a much higher metallicity ($\sim-1.5$) for an old age, both being in stark contrast with the properties of the bulk
of the Eri~II population. However, it has been widely discussed that for a mean metallicity closer to [M/H]= $-2$\footnote{Considering [$\alpha$/Fe] = +0.4, [M/H]= $-2.0$ corresponds to [Fe/H] = $-2.3$.} the observed $\Delta V_{\rm HB}^{\rm \ast bump}$ is smaller that the one predicted by theoretical model. In particular, the value derived here for Eri~II is smaller than predicted by theory by about 0.2~mag, and it is in agreement with previous findings for both globular clusters and dwarf galaxies \citep{Dicecco2010a,Monelli2010c}.

\section{Variable star detection and characterization}\label{sec:vars}

The search for variable stars was performed using a modified version of the Welch \& Stetson index \citep{Welch1993,Stetson1996}, which identifies candidate variable stars on the basis of our multi-band photometry. The light curves of candidate variable stars were individually inspected in order to clean the sample of spurious detections. Given the large time gap between the different data sets, we performed an independent search of periodic variable sources in the ACS and Goodman+DECam data, following the same steps for both samples. For each candidate variable star, we calculated the periodogram of the time series via Fourier analysis \citep[following the prescriptions of][]{Horne1986} exploring the period range between 0.01 and 10~d, which is far broader than the range of all possible periods of RRL stars and AC (which lie between 0.2 and 1.5~d). In order to classify a source as a variable star, we inspected the time series (to verify correlated variations in the different filters) and the morphology of the source in the image (to avoid background galaxies or hot pixels). 

Once the variability was confirmed, the pulsation period was derived and refined by an interactive inspection of the periodogram and the simultaneous visual inspection of the light curves in the six bands. The amplitudes, and intensity-averaged magnitudes for all the variable stars were calculated by fitting the mono-periodic light curves with a set of eight RRL templates based on the set of \citet[][namely RRa1, RRa2, RRa3, RRb1, RRb2, RRb3, RRc, Sine]{Layden1998} plus two templates for Cepheids built by \citet{Bernard2009}. For each folded light curve, the best-fitting template was calculated using the \textit{Simplex} method \citep{Nelder1965}, choosing the template that shows the lowest $\chi^2$. The calculation of the intensity-averaged  magnitudes and amplitudes through the integration of the best-fitting template avoids biases arising from light curves that are not uniformly sampled. 

At the end of the above process, we confirm 69 variable stars in the Goodman+DECam data, 43 of which are also present in the field covered by the ACS data. Only one RRL star (V23) is detected in the ACS but not in the ground-based (DECam+Goodman) data sets, due to the combination of low amplitude and vicinity to a bright star. 67 out of 69 variables are RRL stars, while the remaining 2 stars are ACs. Interestingly, despite the large time baseline and the sparse cadence of the data, no long-period variable stars have been identified. Table~\ref{tab:puls_prop} in Appendix~\ref{ap:puls_param} show the location, periods, intensity-average magnitudes, amplitudes and classification of the 69 variable stars detected in the field of Eri~II. Notes on individual variables are in Table~\ref{tab:comments} (Appendix~\ref{ap:comments}).

We note that the spatial distribution of RRL stars in the outer regions of Eri~II seems highly asymmetric. Indeed, beyond 1.3~r$_{\rm h}$, only 1 RRL star is located on the east side, while 8 are present on the west side and spread over a wide area. It is unlikely that this is an observational bias, as the time sampling and the number of measurements are similar in the two SOAR/Goodman fields. We have verified that the CMD of the stars at distances larger than 1.5~r$_{\rm h}$ is well populated by RGB and HB stars for both sides of the galaxy.

\subsection{Comparison with previous work}\label{sec:stringer}

Using the full six-year data set from the Dark Energy Survey (DES), which covers $\sim$ 5,000~deg$^2$ of the Southern Hemisphere, \citet{Stringer2020} associate five RRL stars to Eri~II. We identify the corresponding stars in our catalogue as V04, V26, V31, V49, and V50. Additionally, the two ACs (V34 and V63) were also detected by \citet{Stringer2020}\footnote{The identifiers of \citet{Stringer2020} for these stars are listed in Table~\ref{tab:comments} (Appendix~\ref{ap:comments}).}. The remaining 62 RRL stars are detected for the first time in this work. We compare the period (\verb+'PERIOD_0'+) given by \citet{Stringer2020} with our periods and most of them agree at the level of 0.0001~d, with the exception of V34 and V49 where the respective discrepancies in period are 0.4182~d and 0.0853~d shorter than those we have measured. We suggest that the period given by \citet{Stringer2020} for V49 is likely an alias since it is close to half of the period we obtain for that star. The discrepancies ($\gtrsim$ 0.1~mag) in some cases among the amplitudes and mean magnitudes between \citet{Stringer2020} and this work are understandable since DES was not designed to study RRL stars. DES only has a few epochs ($\sim$7) per filter but they analysed all bands together in order to detect the RRL stars and determine their periods (see e.g. Figure 3 in \citealt{Stringer2020}). Therefore, the lack of full coverage in the light curve is the likely cause of the differences we observe in the amplitude and magnitudes between the two works.

\section{Anomalous Cepheids}\label{sec:ac}

\begin{figure*}
    \hspace{-0.5cm}
    \includegraphics[width=0.66\textwidth]{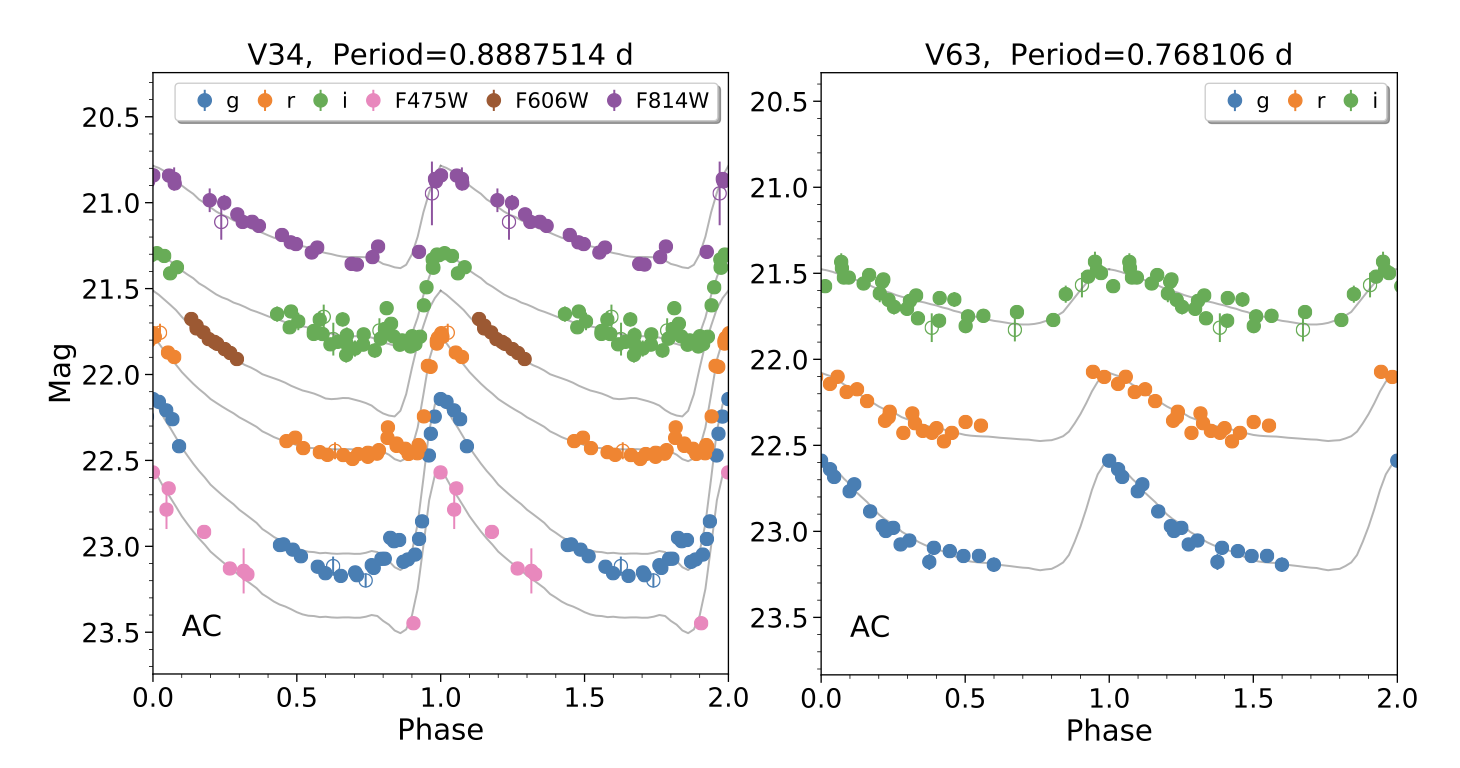}
    \caption{Light curves of the two AC stars detected in Eri~II in the $g, r, i$ (Goodman+DECam) and $F475W, F606W, F814W$ (ACS) bands (when available). The name of the variable and its period are displayed at the top of each panel. Open symbols show the data for which the uncertainties are larger than 3$\sigma$ above the mean error of a given star and, therefore, they were not considered in the period, amplitude and mean magnitude calculations. For clarity, $g, i, F475W, F606W$ and $F814$ light curves have been shifted by +0.6, -0.4, +0.7, -0.2 and -0.6~mag, respectively. Template fits for each light curve are shown by grey lines.}
    \label{fig:lc_ac}
\end{figure*}

Two variable stars (V34, V63) are $\sim$1~mag brighter than the mean magnitude of RRL stars and are classified as ACs. Their light curves are presented in Figure~\ref{fig:lc_ac} and their pulsation properties are in Table~\ref{tab:puls_prop}. One of the two ACs detected is the most distant variable star detected from Eri~II, 5.9\arcmin from its center and relatively far from any other variable star (see Figure \ref{fig:spatial}). The classification of these stars as ACs is based mostly on the location in the CMD, as the period (0.88875~d, 0.76811~d) and the shape of the light curves alone are compatible with a RRab type star. If this was the case, these stars would not be part of Eri~II, but rather a field RRL star accidentally detected along the line of sight. In this scenario, if we derive the distance to this star using the period-luminosity (PL) relation (see \S \ref{sec:distance} for more details), we obtain a distance of $\sim$220 kpc. Finding two MW halo RRL stars at such a distance and clumped in this small area of sky is very unlikely. To quantify this, we integrated the number density profile of RRL stars derived in \citet{Medina2018} and we found that 0.00049 RRL stars are expected in an area of 0.023~deg$^{2}$ (Figure~\ref{fig:spatial}) in the range of distances between 200 and 350~kpc. Therefore, the AC interpretation for these two stars is a more reasonable one.

The classification of the pulsation mode of ACs can be determined from the morphology of well phase-sampled light curves \citep[see e.g., ][]{Soszynski2008b,Plachy2021}. Moreover, theoretical predictions indicate that for the same luminosity and colour, first-overtone (FO) pulsators are less massive than fundamental (F) pulsators \citep[see e.g., ][]{Bono1997b, Marconi2004, Fiorentino2006}. Therefore, determining the correct pulsational mode is important for obtaining a reliable mass estimate. To distinguish the pulsation mode of the two ACs in Eri~II, we follow three approaches. 

Firstly, we visually inspect the folded light curve of both AC stars (see Figure~\ref{fig:lc_ac}). From the $g, r, i$ band, in particular, we can see that V34 has a sawtooth shape, typical of F pulsators, while V63 has a smoother behaviour and sinusoidal shape, typical of a FO type.

\begin{figure}
    \hspace{-0.5cm}
    \includegraphics[width=0.5\textwidth]{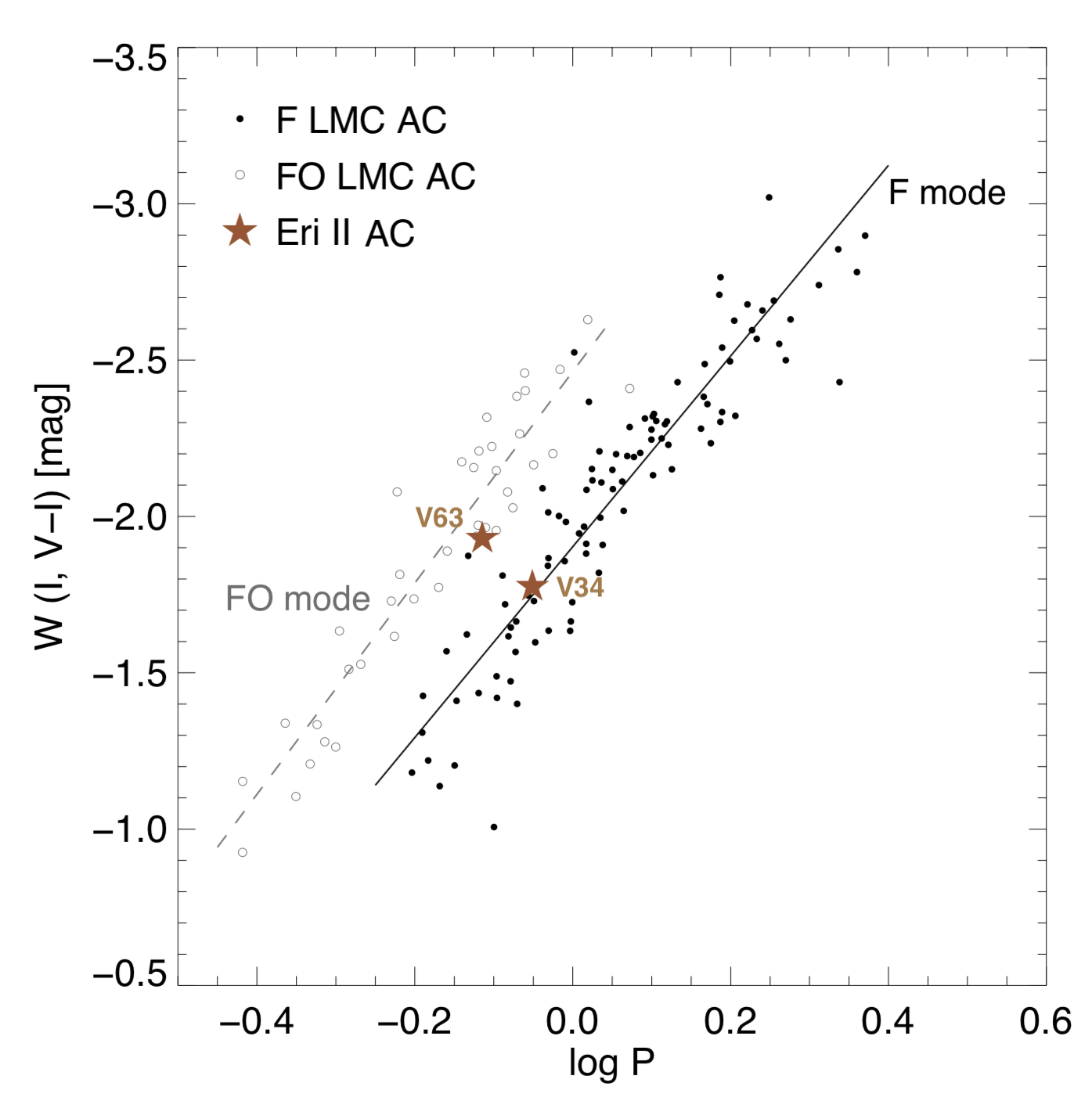}
    \caption{Period-Wesenheit diagram for ACs. Black dots (F pulsators) and grey circles (FO pulsators) represent the ACs of the LMC from the OGLE-IV release \citep[][]{Soszynski2015}. ACs discovered in Eri~II are represented by brown stars. The solid and dashed lines are the empirical PL relations obtained by \citet[][]{Soszynski2015} for ACs in the LMC for the F and FO mode, respectively.}
    \label{fig:ac_pw}
\end{figure}

Secondly, the different pulsation modes follow separate PL relations -- which are also different from those of classical and type II Cepheids. Their location in the PL diagram can therefore be used to constrain their pulsation mode \citep[see, e.g.,][]{Bernard2013}. Figure~\ref{fig:ac_pw} shows that V34 falls on the F mode sequence, while V63 falls closer to the FO mode sequence defined by \citet{Soszynski2015} for the LMC ACs in the period-Wesenheit W($I$, $V-I$) diagram \footnote{In order to plot the ACs in this plane, we convert the $i$ to $I$ using Lupton's photometric transformation equation: \url{http://www.sdss3.org/dr8/algorithms/sdssUBVRITransform.php\#Lupton2005}}. Therefore, based on this method, V34 is clearly a F pulsator while V63 is a FO pulsator.

Thirdly, we used the method described in \citet{Marconi2004} and followed by \citet{Fiorentino2012b} and \citet{MartinezVazquez2016b} to simultaneously constrain the mass and the pulsation mode. It has been shown that ACs obey well-defined mass-dependent period-luminosity-amplitude (MPLA) and period-luminosity-colour (MPLC) relations \citep{Marconi2004}. The MPLA relation is only valid for F pulsators, whereas MPLC relations exist for both pulsation modes. In order to assign a pulsation mode and a mass to each AC, we proceed as follows. We first estimate the mass using both the MPLA and MPLC relations for F pulsators and the MPCL relation alone for FO pulsators. Then, when the two values for the F stars agree with each other within 1-$\sigma$, we classify the star as F mode and take the mean mass as the true value. If instead the two mass estimates are not consistent, we assume that the star is a FO pulsator and we use the corresponding MPLC relation for the mass estimation. In agreement with the classification made from the morphology of the light curves and the period-Wesenheit plane, this method confirms independently that V34 is a F pulsator with M=1.23$\pm$0.06~M$_{\sun}$ and identifies V63 as a FO pulsator with a similar mass, M=1.17$\pm$0.05~M$_{\sun}$ (see Table~\ref{tab:mass_acep}).

\input{mass_acep_tab.tex}

ACs are interesting population tracers, as they can form through two different channels, originating either from the evolution of primordial binaries, or else from the evolution of single, metal-poor (Z $<$ 0.0004), relatively young ($<$ 1~Gyr) stars. \citet{Mateo1995} was the first to notice a correlation between the frequency of ACs (known as $\log S$), normalized to the host galaxy luminosity as a function of its luminosity. This was updated by \citet{Fiorentino2012b}, who suggested that the frequency of ACs is higher in systems with intermediate age population, at fixed magnitude, because of the contribution of the two channels. Despite the fact that the two relations converge for M$_V > $-9~mag, in purely old galaxies such as Eri~II \citep{Simon2021,Gallart2021} only the progeny of binary stars is expected. By assuming a luminosity of M$_V = -7.1$~mag and given the two ACs detected, we derive a frequency of $\log S = 0.52$ for Eri~II, in very good agreement with the general behaviour shown by other galaxies according to the \citet{Mateo1995} correlation.

\begin{figure*}
    \hspace{-0.5cm}
    \includegraphics[width=1.0\textwidth]{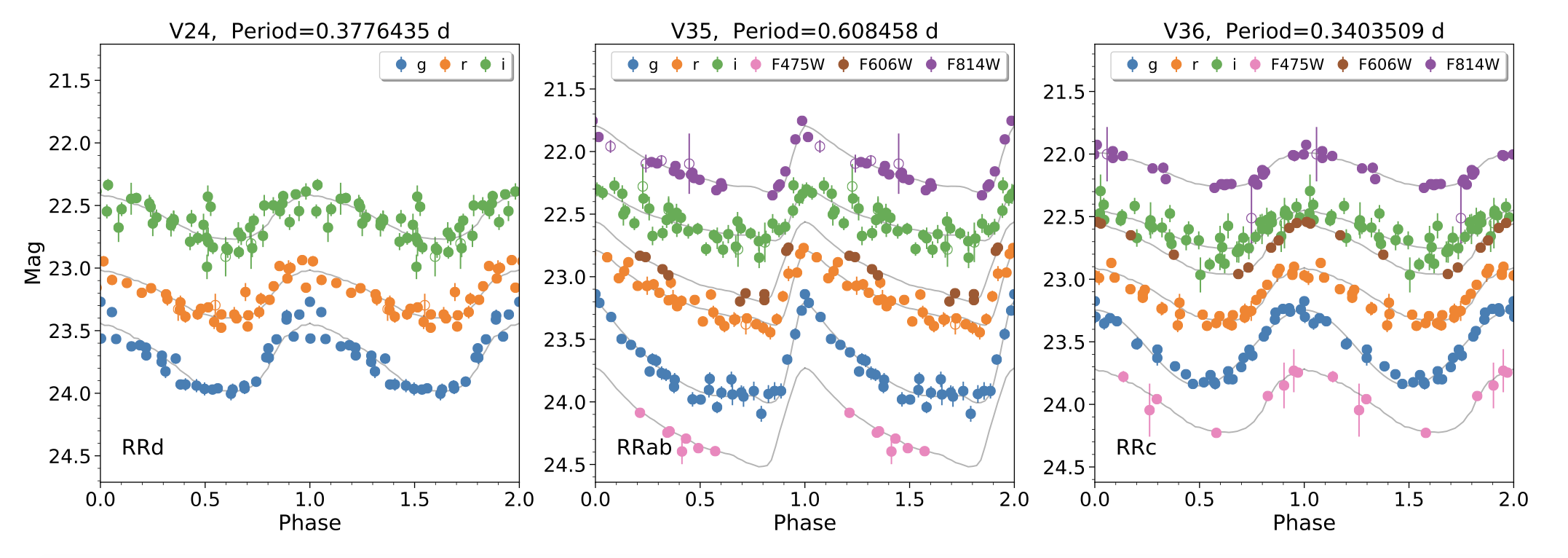}
    \caption{Example of some light curves of the RRL stars discovered in Eri~II in the $g, r, i$ (Goodman+DECam) and $F475W, F606W, F814W$ (ACS) bands (when available). The name of the variable and its period are displayed at the top of each panel. Open symbols show the data for which the uncertainties are larger than 3$\sigma$ above the mean error of a given star and, therefore, they were not considered in the period, amplitude and mean magnitude calculations. For clarity, $g, i, F475W, F606W$ and $F814$ light curves have been shifted by +0.6, -0.4, +0.7, -0.2 and -0.6~mag, respectively. Template fits for each light curve are shown by grey lines. All light curves are available as Supporting Information with the online version of the paper.}
    \label{fig:lc_rr}
\end{figure*}

\input{rrl_summary_tab.tex}

\section{RR Lyrae stars}\label{sec:rrl}

Of the 67 RRL stars found in Eri~II, 44 are fundamental mode (RRab), 6 of them appear to be affected by the Blazhko effect \citep{Blazhko1907}, 12 are first overtone (RRc), 8 are probable double-mode (RRd) pulsators, and 3 are \textit{peculiar} RRL stars (brighter than the bulk of the RRL stars). As explained in \citet[][Section 7.1]{MartinezVazquez2016b}, these \textit{peculiar} variable stars are unlikely to be ACs since they are expected to be brighter, i.e., $\sim$1~mag above the HB in the $r$-band (while they are just between 0.25-0.38~mag brighter). A possible explanation for this apparent overluminosity is that they are either RRL affected by blending, or stars evolving from the blue part of the HB (i.e., BL Herculis variables). However, their amplitude ratios are similar to that those expected for RRL stars (see \S~\ref{sec:amplitudes}) thus not supporting a possible blend effect\footnote{In addition, a visual inspection of each star in the images does not reveal a source nearby.}. On the other hand, BL Herculis variables typically have periods longer than 1 day \citep[see e.g.,][]{Soszynski2010} but these peculiar variables have periods lower than 0.8~d. Variable stars of this kind have been detected in galaxies like Sculptor \citep{MartinezVazquez2016b}, Carina \citep{Coppola2013}, also Cetus and Tucana \citep{Bernard2009}.\\
The spatial distribution of the different types of RRL stars and their location in the CMD are displayed in Figure~\ref{fig:spatial} and ~\ref{fig:cmd} with different symbols and colours. The individual pulsation properties together with their location and classification are given in Table~\ref{tab:puls_prop}. Figure~\ref{fig:lc_rr} shows the light curves in the $g$ (blue), $r$ (red), $i$ (yellow), $F475W$ (purple), $F606$ (green) and $F814W$ (grey) pass-bands for six RRL stars. The full set of light curves in the ACS and DECam filters for the 41 RRL stars in ACS and the 66 RRL stars in Goodman+DECam is available as Support Information Material in the online version of the journal.

\subsection{Deepening the Oosterhoff classification}

\begin{figure}
    \centering
    \includegraphics[width=0.5\textwidth]{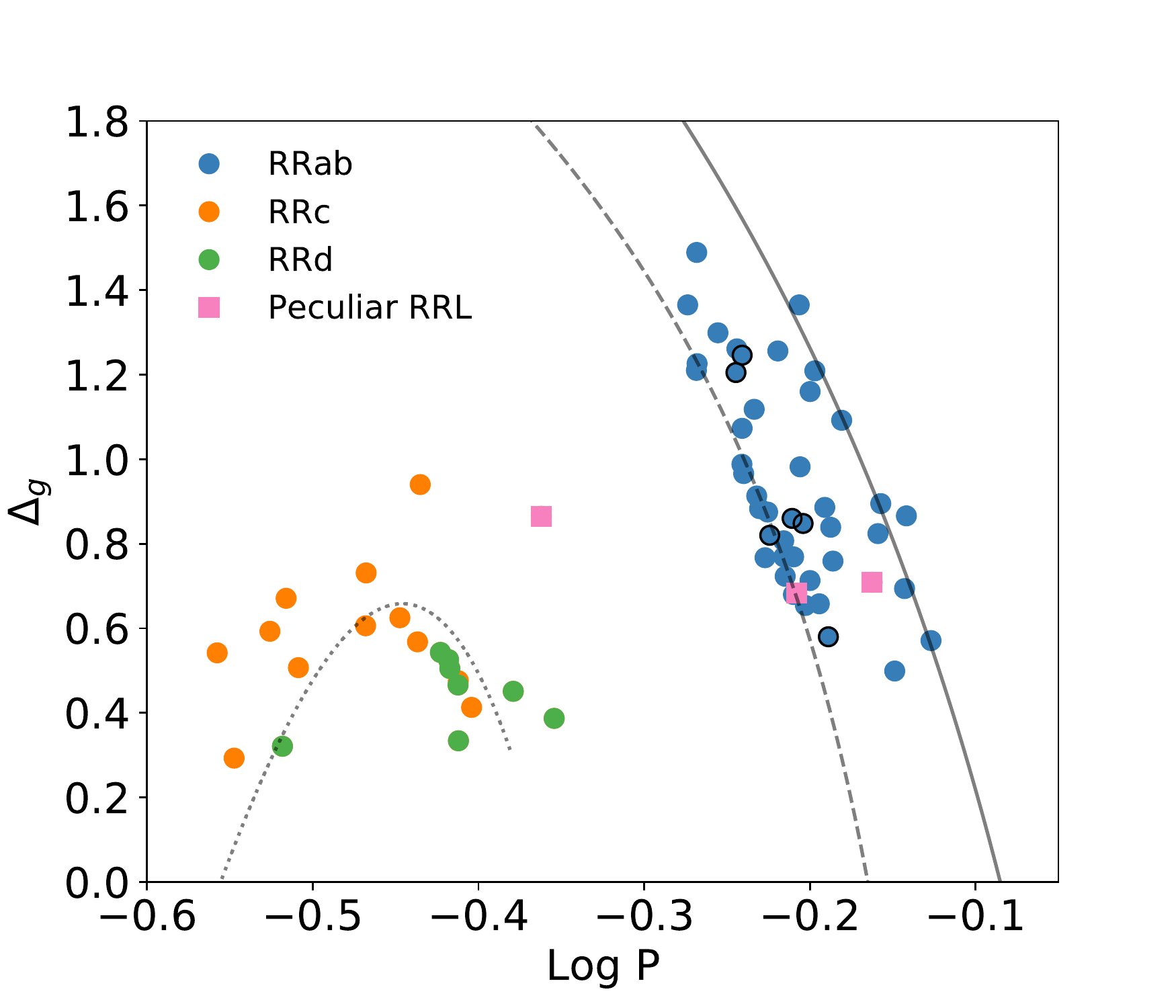}
    \caption{Bailey diagram in the $g$-band for the RRL stars of Eri~II. The dashed and solid lines are the relations  for the RRab stars in Oosterhoff I and Oosterhoof II type globular clusters, respectively. The dotted curve is that derived empirically for the RRc stars in the cluster M22 (Oo-II type). Black circles point to those RRab stars we consider affected by the Blazkho effect.}
    \label{fig:bailey}
\end{figure}

Table~\ref{tab:summary_rrl} reports the number, the mean magnitude in the different filters, and the mean period of the different types of pulsators. We find that the frequency of RRc and RRcd types are f$_c = \frac{Nc}{Nab+Nc} = 0.21$ and f$_{cd} = \frac{Nc+Nd}{Nab+Nc+Nd} = 0.31$, respectively. By comparing with the values obtained in other dwarf galaxies, we find the Eri~II result is similar to the average expected in dwarf galaxies. In particular, MW dwarf satellites like Ursa Major~I, Hercules, Sextans, Fornax, LMC; M31 dwarf satellites like And~XI, And~XV, And~VII, NGC~147, NGC~185; and isolated galaxies like Leo~A and IC~1613, share almost the same frequency of RRcd stars,  f$_{cd} \sim 0.3$ (see Table 6 in \citealt{MartinezVazquez2017}). The mean periods obtained here for the RRab and the RRc are 0.62 and 0.34 d. If we compare these values with those driving the Oosterhoff classification of globular clusters \citep{Oosterhoff1939, Oosterhoff1944}, Eri~II corresponds to a system with an Oosterhoff intermediate (Oo-int) classification \citep[see, e.g., ][]{Bono1994, Smith1995}, which is the usual classification given for dwarf galaxies \citep{Catelan2009}.

Figure~\ref{fig:bailey} presents the period-amplitude (or Bailey) diagram. The dashed and solid lines show the loci provided by \citet{Fabrizio2019}, scaled to the $g$ band applying a factor of 1.29 provided by \citealt{Vivas2020a}, for the RRab stars in Oosterhoff I (Oo-I) and Oosterhoof II (Oo-II) type globular clusters. The dotted curve is that derived by \citet{Kunder2013} for the RRc stars in M~22, an Oo-II type cluster. Despite the fact that the mean period of RRab type, $\langle P \rangle$=0.62~d would suggest an Oo-int type galaxy, the distribution in the period-amplitude plane clearly shows that while the majority of RRab type stars (66\%) closely follow the Oo-I line, 34\% of them are scattered in between the two lines or closer to the Oo-II locus. Dwarf galaxies like Ursa Major~I, Sextans, Leo~II, And~XIII, And~XVI, And~XXV, NGC~185, Tucana, IC~1613, and ESO410-G005 have essentially the same distribution in the Bailey diagram as Eri~II (see \%Oo-I and \%Oo-II values of Table 6 in \citealt{MartinezVazquez2017}). Further, in \citet{MartinezVazquez2017} we find that the vast majority of galaxies host a large fraction of RRab that match the Oo-I locus, between 60\% and 90\% despite having mean periods that would classify them as Oo-int systems. This suggests once more that the RRL stars in complex systems such as galaxies are not properly represented by the mean period of their RRab stars. Hence, the Oosterhoff classification based on the mean period poorly describes the complicated early history of dwarf galaxies.

\subsection{Amplitude ratios in DECam filters}\label{sec:amplitudes}

\begin{figure*}
    \centering
    \includegraphics[width=1.0\textwidth]{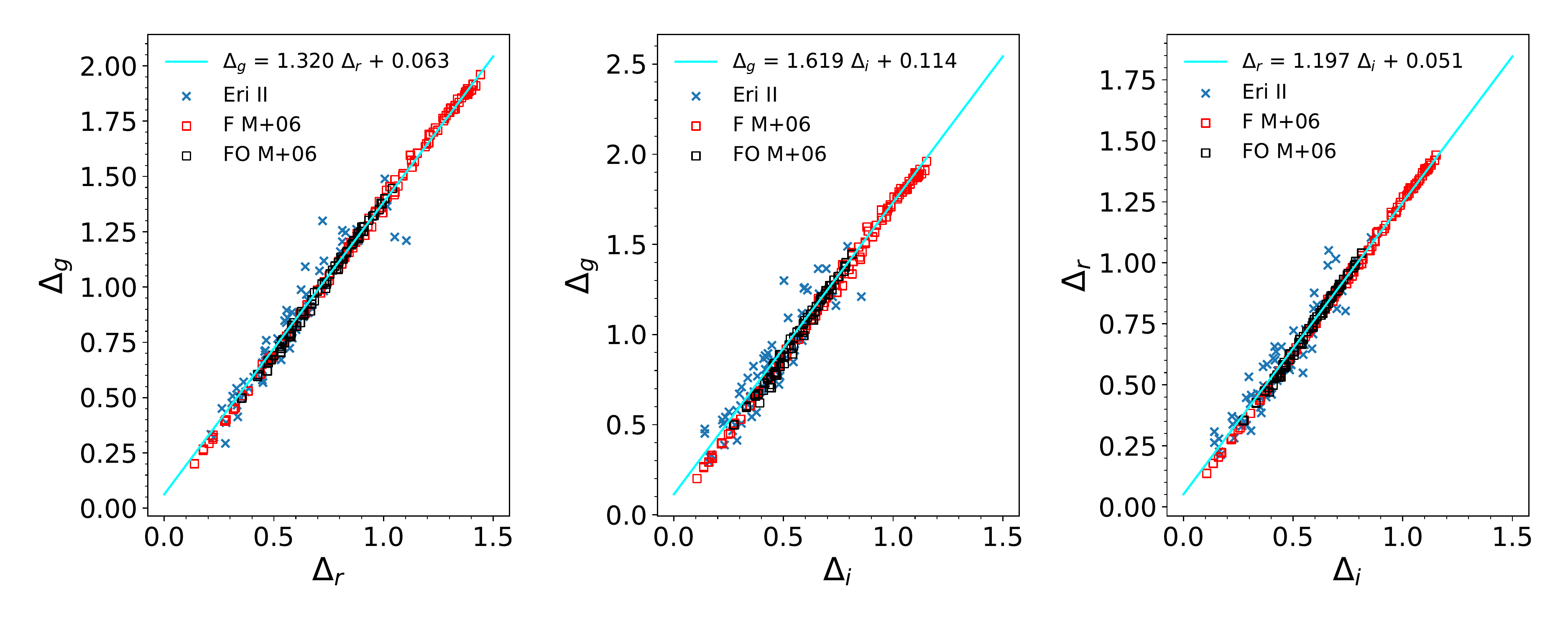}
    \vspace{-1.0cm}
    \caption{Amplitude ratios for the RRL stars in Eri~II (crosses). The cyan line represents the fit to the data. For comparison we show with black and red squares the fundamental and first overtone amplitudes predicted theoretically for different populations in \citet{Marconi2006}.}
    \label{fig:amplitude_ratios}
\end{figure*}

We have also derived the amplitude ratios for the three combinations of $g$, $r$, ad $i$ bands. These relations are useful for future comparisons and transformations between SDSS-like filters, such as DECam and the forthcoming filters with which the Vera C. Rubin Observatory Legacy Survey of Space and Time (LSST) will be carried out \citep{Ivezic2019}. Figure~\ref{fig:amplitude_ratios} shows from the left to the right, $\Delta_g$ \textit{vs.} $\Delta_r$, $\Delta_g$ \textit{vs.} $\Delta_i$, $\Delta_r$ \textit{vs.} $\Delta_i$. The Eri~II RRL stars, represented in Figure~\ref{fig:amplitude_ratios} by blue crosses, are used to derive a linear fit (cyan solid lines):
\begin{eqnarray}
\Delta_g = 1.32 (\pm 0.05) \Delta_r + 0.06 (\pm 0.03) ~ ~ ~ ~ \sigma = 0.08 \label{eq:AgAr} \\
\Delta_g = 1.62 (\pm 0.08) \Delta_i + 0.11 (\pm 0.04) ~ ~ ~ ~ \sigma = 0.11 \label{eq:AgAi} \\
\Delta_r = 1.20 (\pm 0.05) \Delta_i + 0.05 (\pm 0.02) ~ ~ ~ ~ \sigma = 0.07 \label{eq:ArAi}
\end{eqnarray}

We compare with the predicted theoretical pulsation amplitudes for the full set of F and FO models derived by \citet{Marconi2006} for RRL stars in the SDSS photometric system (red empty squares in Figure~\ref{fig:amplitude_ratios}). We see clearly in Figure~\ref{fig:amplitude_ratios} that the the amplitude ratios obtained using the RRL stars in Eri~II are in good agreement with the theoretically predicted values of \citet{Marconi2006}. 

\subsection{Distance determination}\label{sec:distance}

RRL stars are fundamental distance indicators and play a crucial role in establishing the population II distance scale \citep{Beaton2018}. RRL stars have been routinely used to derived the distance to nearby galaxies. In this way, using different methods and different calibrations MW ultra-faint \citep[e.g.][]{Vivas2016a, MartinezVazquez2019,Vivas2020b} and classical dwarf galaxies \citep[e.g.][]{MartinezVazquez2015,Coppola2015, Vivas2019}, M31 dwarf galaxies \citep[e.g.][]{Fiorentino2012c, MartinezVazquez2017, Monelli2017}, isolated dwarf galaxies \citep[e.g.][]{Bernard2009,Bernard2010} and Sculptor group dwarf galaxies \citep[e.g.][and references therein]{Yang2014} have RRL distance determinations.

In order to determine the distance to Eri~II from its RRL stars, we used the theoretical PL relation derived by \citet{Caceres&Catelan2008} in the $i$-SDSS band, the PL relation in the $I$-Cousins band from \citet{Marconi2015}, and the PL relation in the $F814W$ band obtained following the same approach as in \citet{Marconi2015}.

We transform our $i$ band into $i$-SDSS and into $I$-Cousins, using the transformation equations provided by the DES Collaboration\footnote{\url{http://www.ctio.noao.edu/noao/content/Photometric-Standard-Stars-0\#transformations}} and Lupton\footnote{\url{http://www.sdss3.org/dr8/algorithms/sdssUBVRITransform.php\#Lupton2005}} so we can use some of the previously mentioned PL relations. In addition, we correct our photometry for extinction by considering a reddening of E($B-V$) = 0.01 \citep{Schlegel:1998} and the following extinction coefficients, $A_{i}/A_{V} = 0.51$ \citep[obtained from][]{DES_DR1}, $A_{F814}/A_{V} = 0.48$, and $A_{I}/A_{V} = 0.58$ \citep[derived using][]{Cardelli1989}. In order to provide an accurate and precise distance we remove the RRd stars and those RRL stars with the \textit{noisy} qualifier (based on their light curve) in Table~\ref{tab:comments}. We end up with a sample of 46 RRL stars in the Goodman+DECam photometry and 30 in the ACS photometry.

\input{distance_moduli_tab.tex}

The mean true distance moduli obtained with the different PL relations using the selected clean sample of RRab and RRc stars, assuming [Fe/H] = $-2.38$ \citep{Li2017}, are listed in Table~\ref{tab:distance}. A 3$\sigma$ clipping is applied in these calculations. Only 1 RRL star in the ACS photometry is rejected by this procedure. The systematic uncertainties in the distance moduli are derived by propagation of errors considering: \textit{(i)} the photometric uncertainties of the mean magnitudes in $i$ ($\sim$ 0.07~mag) and $F814W$ ($\sim$ 0.03~mag); \textit{(ii)} the uncertainty that comes from transforming $i$ to $i$-SDSS and $I$, when it applies; \textit{(iii)} the uncertainties of the coefficients in the relationships; \textit{(iv)} uncertainties of 0.2~dex in [Fe/H], 0.2 dex in [$\alpha$/Fe] (when it applies), and 0.0001~d in the period; \textit{(v)} the E($B-V$) uncertainty, assumed to be 10\%. The random uncertainties are quite small in all cases (0.01~mag). They are estimated from the standard error of the mean (the standard deviation divided by the square root of the number of RRL stars used to estimate the distance modulus).  

All the obtained distances agree at 0.2$\sigma$ level. We consider therefore the final distance modulus of Eri~II as the average of them: $\mu_0$ = 22.84 $\pm$ 0.05 mag ($\sigma$ = 0.01~mag). This corresponds to a heliocentric distance for Eri~II of $D_{\sun} = 370 \pm 9$~kpc. 

The distance modulus derived for Eri~II using its population of RRL stars is consistent within 1$\sigma$ to that found by \citet[][$\mu_0 = 22.8 \pm 0.1$ mag]{Crnojevic2016b} and \citet[][$\mu_0 = 22.9 \pm 0.1$ mag]{Koposov2015}, however it is 2$\sigma$ larger than the distance moduli derived by \citet[][$\mu_0 = 22.6$ mag]{Bechtol2015} and \citet[][$\mu_0 = 22.65 \pm 0.08$ mag]{Simon2021}. Such a discrepancy with \citet{Bechtol2015} can be probably explained by their shallow photometry, while the origin is not clear in the case of \citet{Simon2021}.

\subsection{Period-luminosity relations in DECam filters}\label{sec:pls}

\begin{figure*}
    \centering
    \includegraphics[width=0.33\textwidth]{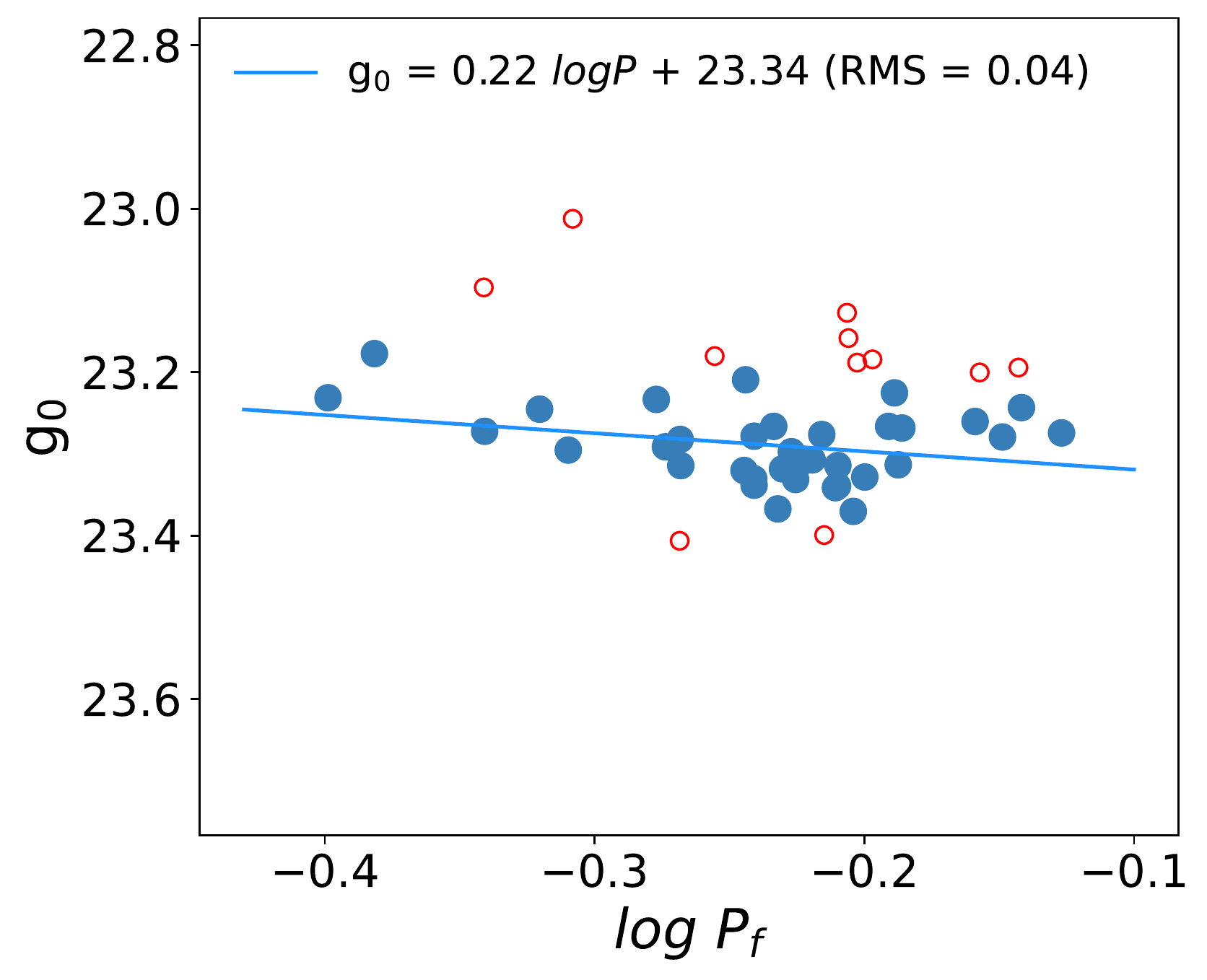}
    \includegraphics[width=0.33\textwidth]{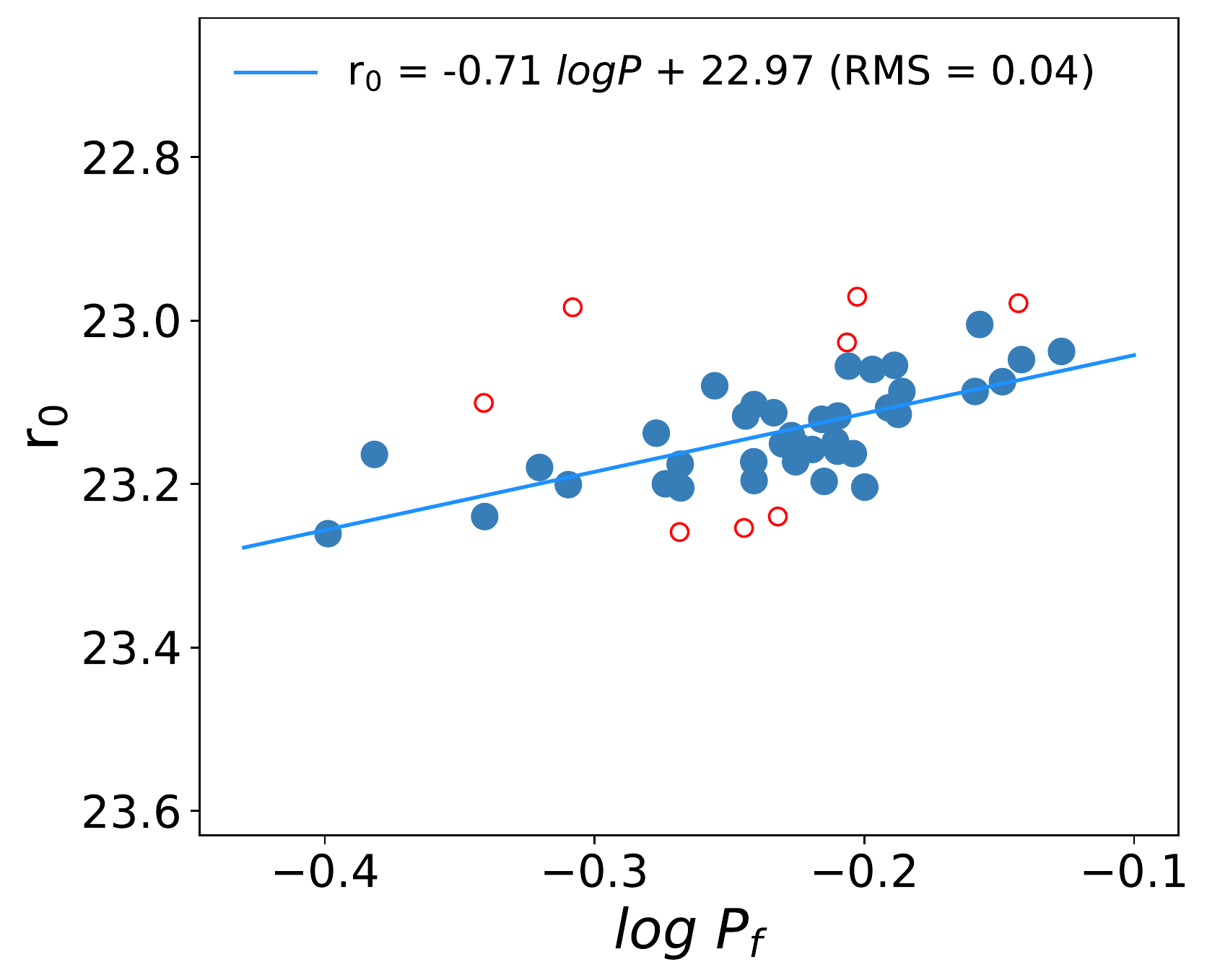}
    \includegraphics[width=0.33\textwidth]{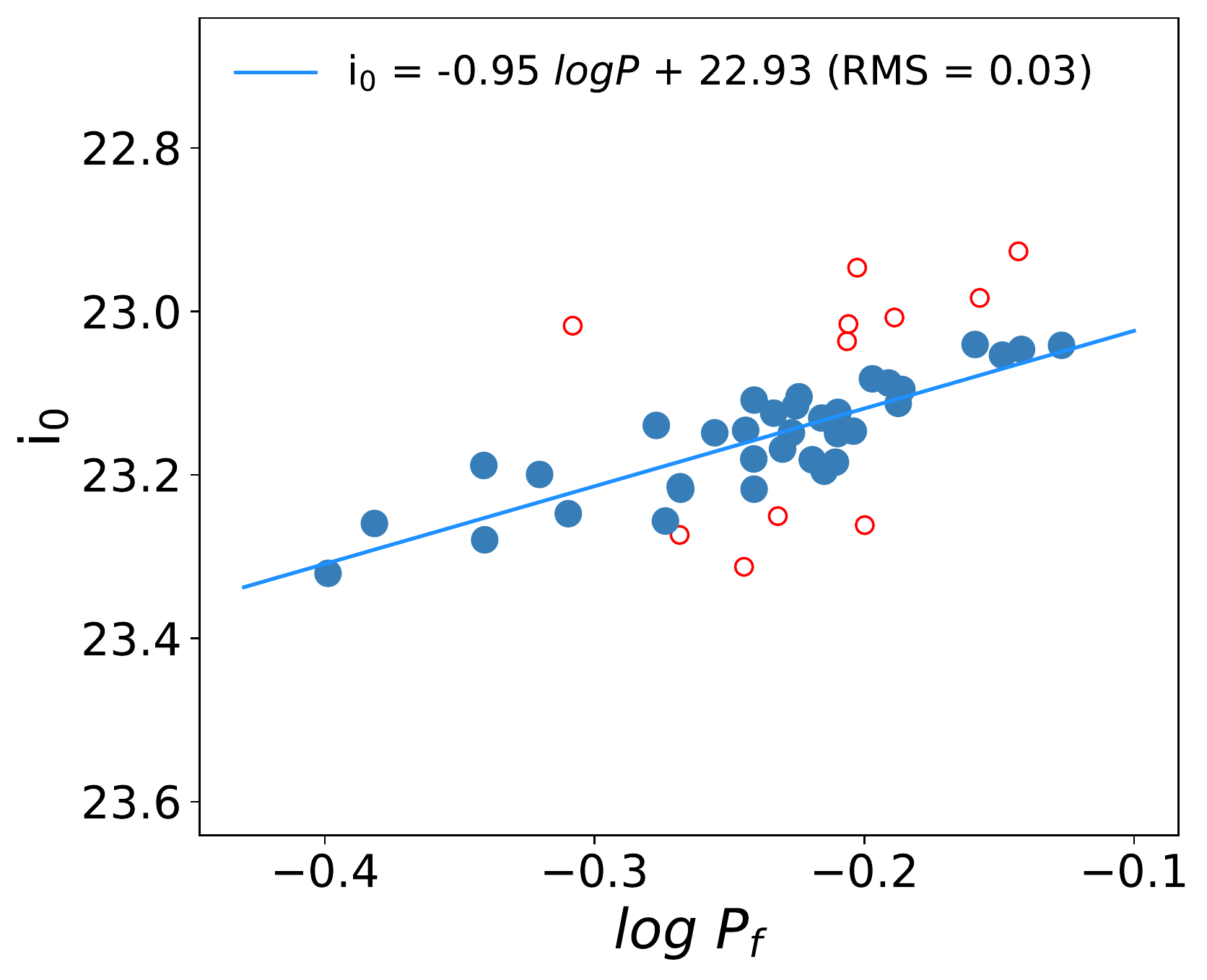}
    \caption{Period-luminosity relations for the RRL stars in Eri~II. From
    left to right, the dereddened $g_0$, $r_0$, and $i_0$ magnitudes are shown as a 
    function of the logarithm of the period. Only stars represented with blue 
    filled circles have been used in the fit, while stars shown as open red circles
    have been rejected by an iterative 2$\sigma$-clipping algorithm. }
    \label{fig:pls}
\end{figure*}

We have derived the following period luminosity relations in the $g,r,i$ bands for the RRL stars of Eri~II, using a clean sample of stars (see Figure~\ref{fig:pls}): 
\begin{eqnarray}
g_0 = 0.22 (\pm 0.12) \log P_{f} + 23.34 (\pm 0.03) ~ ~ ~ ~ \sigma = 0.04 \label{eq:plg_eri2} \\
r_0 = -0.71 (\pm 0.11) \log P_{f} + 22.97 (\pm 0.03) ~ ~ ~ ~ \sigma = 0.04 \label{eq:plr_eri2} \\
i_0 = -0.95 (\pm 0.10) \log P_{f} + 22.93 (\pm 0.02) ~ ~ ~ ~ \sigma = 0.03 \label{eq:pli_eri2}
\end{eqnarray}

\noindent where P$_{f}$ is the fudamentalized period of the RRL stars, i.e., $P_{f} = P$ for RRab stars and $P_{f} = 10^{\log P + 0.127}$ for RRc stars \citep[e.g., see][]{Bono2001}. In order to determine these relationships, we dereddened the apparent magnitudes $g,r,i$ of Eri~II RRL stars using a reddening E($B-V$) = 0.01 mag and we used an extinction law corresponding to values of $R_g = 3.186$, $R_r = 2.140$, $R_i = 1.569$ \citep{DES_DR1}. 

Using the true distance modulus derived in this work, we scale the intercept of the above relationships to obtain the absolute PL relations in the $g,r,i$ bands:
\begin{eqnarray}
M_g = 0.22 (\pm 0.12) \log P_{f} + 0.50 (\pm 0.06) ~ ~ ~ ~ \sigma = 0.04 \label{eq:aplg_eri2} \\
Mr = -0.71 (\pm 0.11) \log P_{f} + 0.13 (\pm 0.06) ~ ~ ~ ~ \sigma = 0.04 \label{eq:aplr_eri2} \\
M_i = -0.95 (\pm 0.10) \log P_{f} + 0.09 (\pm 0.05) ~ ~ ~ ~ \sigma = 0.03 \label{eq:apli_eri2}
\end{eqnarray}

The intercepts are in agreement with those obtained in the DECam bands by \citet{Vivas2017} for the globular cluster M5, however the slopes here differ to those obtained in that work. The cause of this discrepancy may be due to a different data treatment and/or the different samples fitted.

In particular, if we compare the PL($i$) obtained for Eri~II (eq.~\ref{eq:apli_eri2}) with the theoretical PL($i$-SDSS) predicted by \citet{Caceres&Catelan2008} assuming a [Fe/H] = $-2.38$ and [$\alpha$/Fe] = +0.4 (i.e, $M_i = 1.035 \log P + 0.061$), we can see $\sim 1\sigma$ agreement between them.

For the $g$ and $r$ band, there are not theoretical relations to compare with, however we see that their period slopes seems to follow the expected trend \citep[positive for $g$ as in the $B$ band, null for the $V$ band, and becoming more negative for redder filters; see e.g. Fig 6 in][]{Beaton2018}.

These are the first PL relations obtained for a low metallicity system in the DECam filters and could be used in systems with approximately the same metallicity range as Eri~II, i.e., in metal-poor systems with [Fe/H] $< -2.0$ dex.

\subsection{Metallicity distribution of RR Lyrae stars}\label{sec:metal}

\begin{figure}
    \centering
    \includegraphics[width=0.5\textwidth]{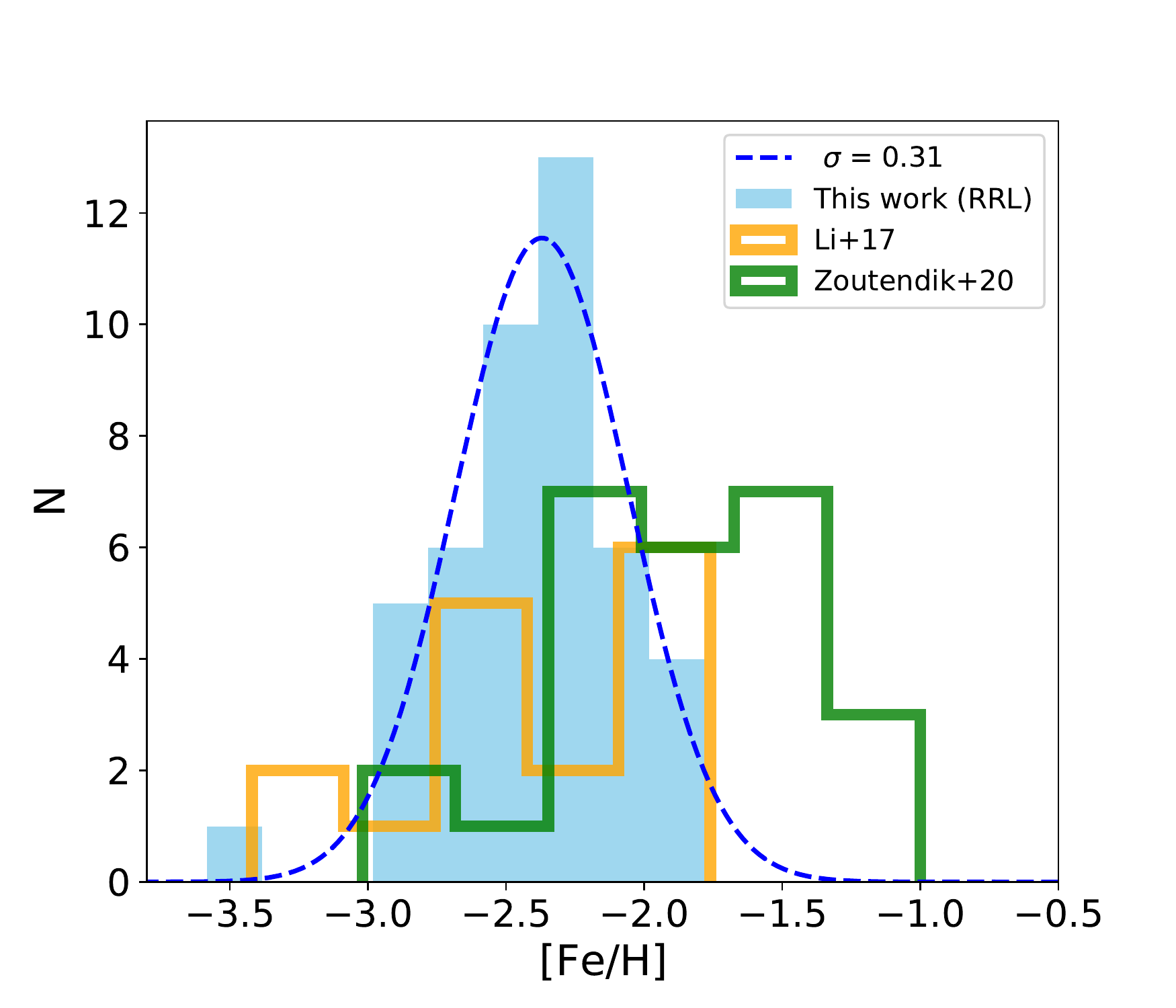}
    \caption{Metallicity distribution of Eri~II from RRL stars obtained in this work (blue histogram) compared to spectroscopic measurements for RGB stars available in the literature (orange, \citealt{Li2017}; green, \citealt{Zoutendijk2020}).}
    \label{fig:FeH_hist}
\end{figure}

Following \citet{MartinezVazquez2016a} we derived the metallicity distribution for the RRL stars using the PL relations in the $i$-SDSS band from \citet{Caceres&Catelan2008}, and the $I$ and $F814W$ bands based on the models by \citet{Marconi2015}. The average metallicity distribution obtained for the clean sample of RRL stars using the latter relations is displayed in Figure~\ref{fig:FeH_hist}. By construction, the mean values of our metallicity distributions match with the mean [Fe/H] of Eri~II given by \citet{Li2017}, since we made use of this value to obtain the distance modulus. The dispersion of this distribution, assessed using a Gaussian fit is $\sigma_{\rm [Fe/H], RRL}$ = 0.3~dex.

As discussed in \citet{MartinezVazquez2016a}, the metallicity distribution based on well sampled light curves of a monometallic population such as that of a normal globular cluster like Reticulum provides a metallicity dispersion of $\sigma_{\rm [Fe/H], RRL} \simeq $  0.2~dex. Considering this value as an estimate of the dispersion of the method, a measured dispersions of 0.31~dex from the RRL stars in Eri~II is consistent with an intrinsic metallicity spread of $\simeq$ 0.2~dex being in place at a very early epoch. This implies a fast chemical evolution, in agreement with both spectroscopic measurements \citep{Li2017} and the SFH derivation \citep{Simon2021,Gallart2021}.

Figure~\ref{fig:FeH_hist} also shows the comparison with spectroscopic determinations; in orange, the metallicity distribution derived for 15 RGB stars by \citet{Li2017}, while in green is the distribution derived by \citet{Zoutendijk2020} and based on MUSE spectra for 26 RGB stars. Our RRL distribution, which represents the global old population of Eri~II, looks more peaked than the other two, but this could be at least partly due to the smaller sample of spectroscopic stars available. The dispersion of the metallicity distribution derived from the RRL stars of Eri~II is slightly smaller than that shown by the spectroscopic studies. This could be at least partly explained by taking into account that the RGB samples could include slightly younger stars, whose existence can not be completely excluded by the SFH determinations \citep[detected by][]{Simon2021,Gallart2021}, for which there is no counterpart in the RRL group. In addition, it has to be taken into account that the most metal-poor and most metal-rich tails of the distribution are probably not producing RRL stars because these stars do not fall into the instability strip. Finally, we should not discard the possibility that field contamination may be present in the MUSE data set. 

Given our sample of Eri~II RRL stars are all at the same distance, we can exclude the possibility that the sample is contaminated by field stars. Therefore, we conclude that the metallicity distribution obtained in this work adequately describes the old and metal-poor population of Eri~II. 

\begin{figure}
\centering
\hspace{-1.0cm }
    \includegraphics[width=0.50\textwidth]{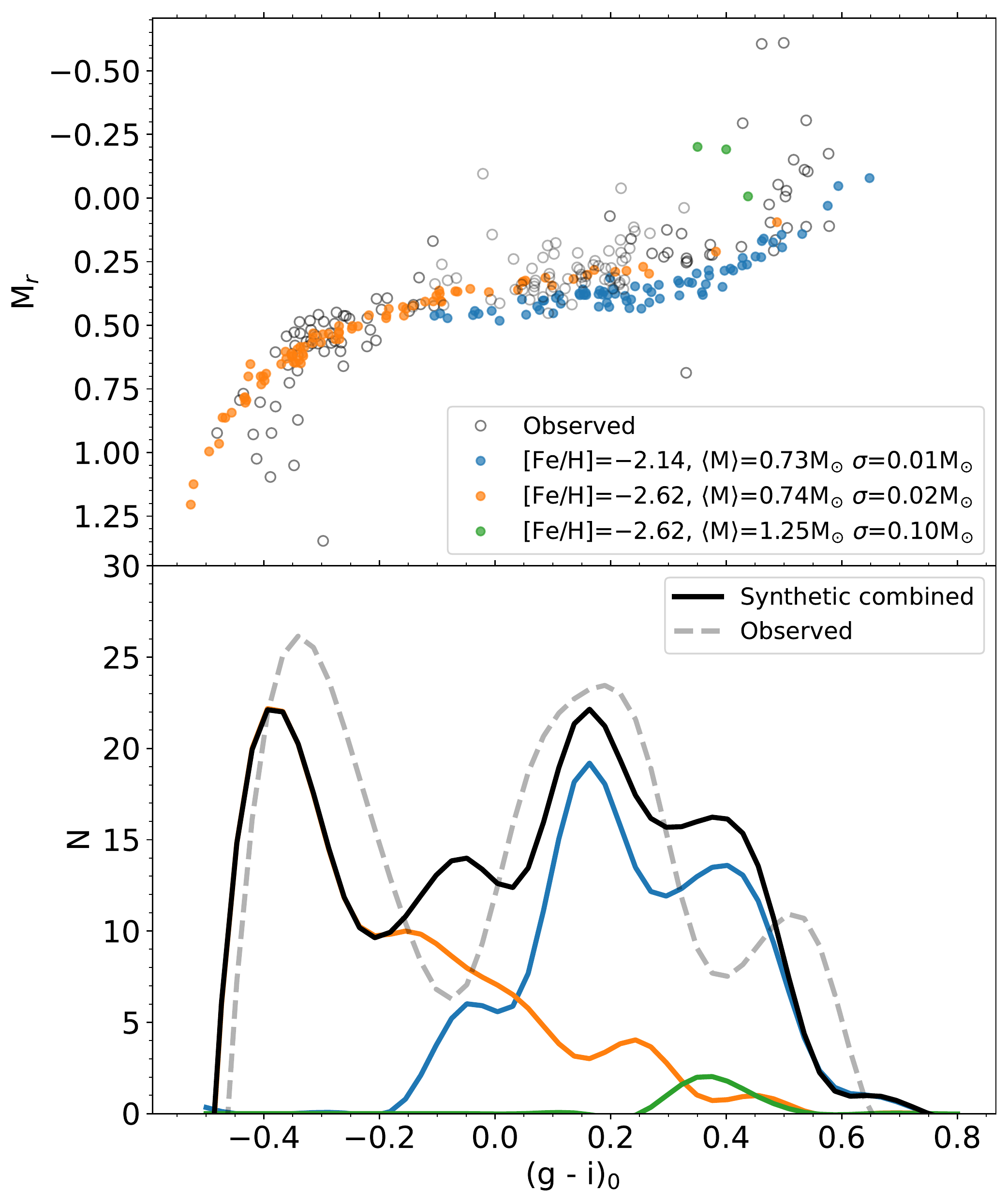} 
    \caption{\textit{Top}. Absolute CMD (M$_r$, $(g-i)_0$) for the observed and the simulated HB of Eri~II. The constraints for each synthetic populations are labeled in the legend. \textit{Bottom}. Colour $(g-i)_0$ distribution of each synthetic population, the synthetic combined, and the observed HB population. The synthetic HB accounts for photometric errors based on Monte Carlo simulations.}
    \label{fig:hbsimul}
\end{figure}


\begin{figure}
\hspace{-0.8cm}
    \includegraphics[width=0.50\textwidth]{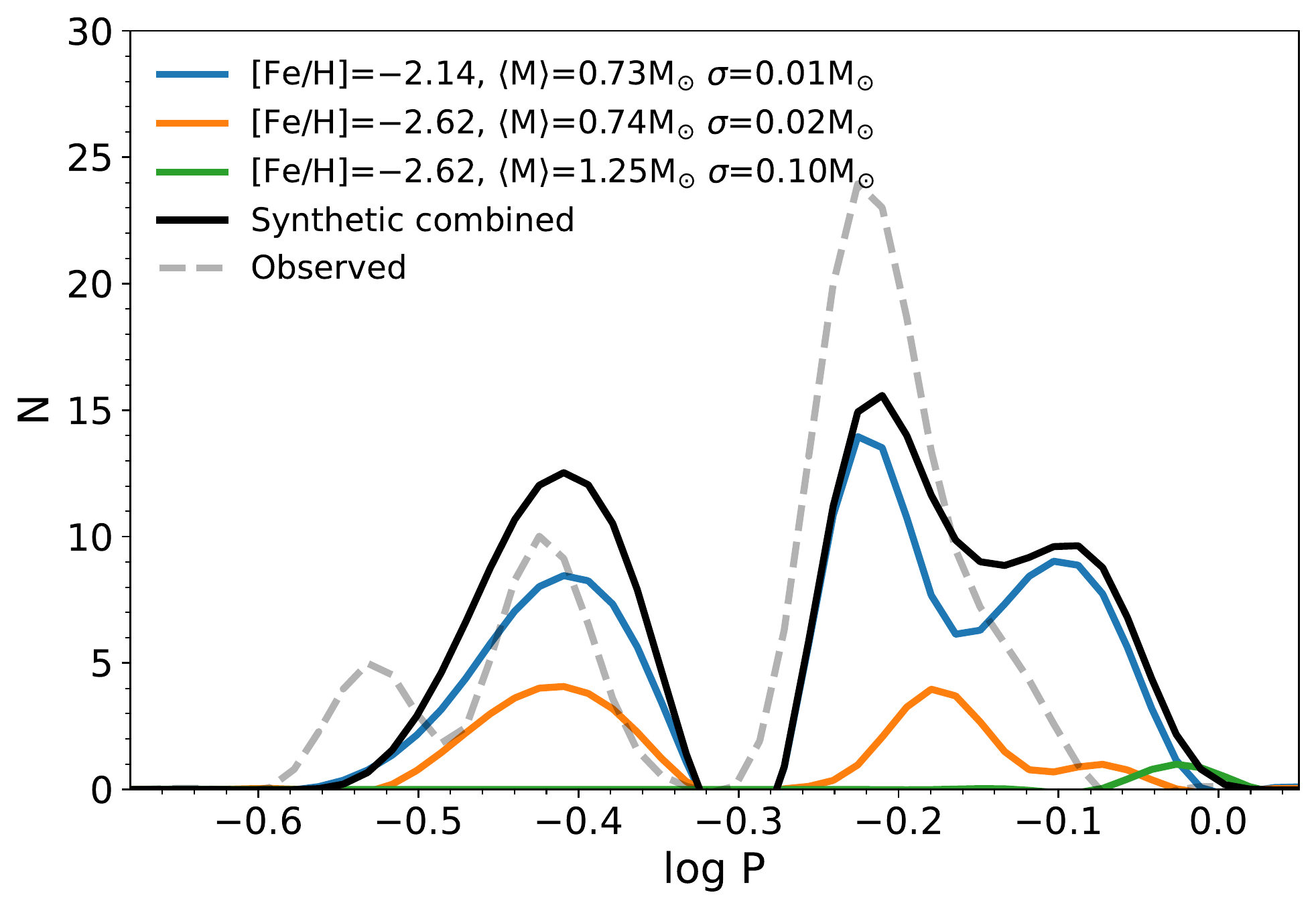} 
    \caption{Period distribution of the observed and simulated RRL stars.}
    \label{fig:persimul}
\end{figure}

\section{Horizontal branch simulations}\label{sec:simul}

In order to further search for evidence of a significant metallicity spread among the various stellar populations of Eri~II we investigated in some detail the distribution of stars in the observed HB by using synthetic HB simulations. Synthetic HB models have been computed by adopting the HB tracks from the BaSTI stellar models repository \citep{basti:04, basti:06}. Using the $\alpha$-enhanced BaSTI models we selected the model sets corresponding to Z=0.0001 ([Fe/H]$=-2.62$) and Z=0.0003 ([Fe/H]$=-2.14$) based on the metallicity distribution obtained in \S~\ref{sec:metal}. The numerical code used to generate the synthetic HB distribution has been developed in the framework of the BaSTI project, and a  description can be found in \citet{cs:13} and in \citet{dalessandro:13}. Since in the case of Eri~II we are considering only quite old stellar populations \citep{Simon2021,Gallart2021}, our synthetic HB calculations need, for each individual stellar population, the specification of three parameters, \textit{(i)} the metallicity, \textit{(ii)} the mean mass $\langle M \rangle$ of the stars along the HB, and \textit{(iii)} the mass spread $\sigma \langle M \rangle$ around this mean value. Throughout the present analysis, a Gaussian distribution around the mean mass value has been assumed, and $\sigma \langle M \rangle$ represents the $1\sigma$ spread around the specified mean mass value. 

To reproduce the observed HB morphology we combined the synthetic HB models corresponding to three individual  populations: 1) a very metal-poor, old stellar population with Z=0.0001, 2) a more metal-rich, old population with Z=0.0003, and 3) a (very small) sample of HB stars corresponding to the blue straggler stars progeny, that we assumed to belong to the more metal-poor stellar population\footnote{The results of our HB simulations should be barely affected by a different assumption about the metallicity of the bulk of the blue straggler stars population. This is because in the metallicity range covered by the stellar populations in Eri~II, HB stellar models more massive than about $1.1M\odot$ occupy the same region of the CMD \citep[see, e.g.][and references therein]{sc:05}.}

We have performed several HB simulations by taking care that the total number of synthetic HB stars is the same as the observed number. The observational constraints that we used to find the best synthetic HB simulation are the HB colour distribution and the RRL period distribution; which are strongly affected by both the mass and metallicity distribution adopted in the synthetic HB models.

Figure~\ref{fig:hbsimul} shows the comparison of the observed HB and the simulated one that closely reproduces the colour distribution. The top panel shows in black the observed points, while the orange, blue, and green symbols are used for three different synthetic populations, according to the labelled values of metallicity, mean mass, and mass dispersion. The bottom panel shows the colour distribution: the black solid and dashed line represents the observed and the sum of the synthetic populations, respectively. Although qualitative, this comparison shows that a metallicity spread is required to account for the colour range covered by the Eri~II HB stars, in excellent agreement with the results obtained from the SFH by \citet{Gallart2021} and for the RRL stars (see \S~\ref{sec:metal}).

Figure~\ref{fig:persimul} shows the comparison between the observed RRL period distribution and that corresponding to the variable stars in the synthetic HB model shown in Figure~\ref{fig:hbsimul}. This figure reveals that the selected HB simulation adequately reproduces the location of the two main peaks in the observed period distribution, although the simulation predicts the presence of a (small) sample of long-period RRL stars that are not observed. To assume a different average age for the metal-rich population would not help to remove these long period variables from the HB simulation; this is because in order to achieve a significant reduction of the luminosity, and hence of the period, in the corresponding RRL population, it would  be necessary to assume a huge reduction -- of the order of 5-6~Gyr -- for the age of the bulk metal-rich population, a possibility ruled out by the accurate SFH derived by \citet{Gallart2021}. A more realistic option to overcome this problem would be to adopt a multimodal mass distribution for the metal-rich stellar population, but this would need the introduction of two additional free parameters. When considering the small number of objects populating this portion of the period distribution, and the aim of this analysis, a more complex HB simulation appears unprofitable.

\section{Metallicity gradients and radial distribution of RR Lyrae stars}\label{sec:discussion}

\begin{figure}
    \hspace{-0.6cm}
    \includegraphics[width=0.49\textwidth]{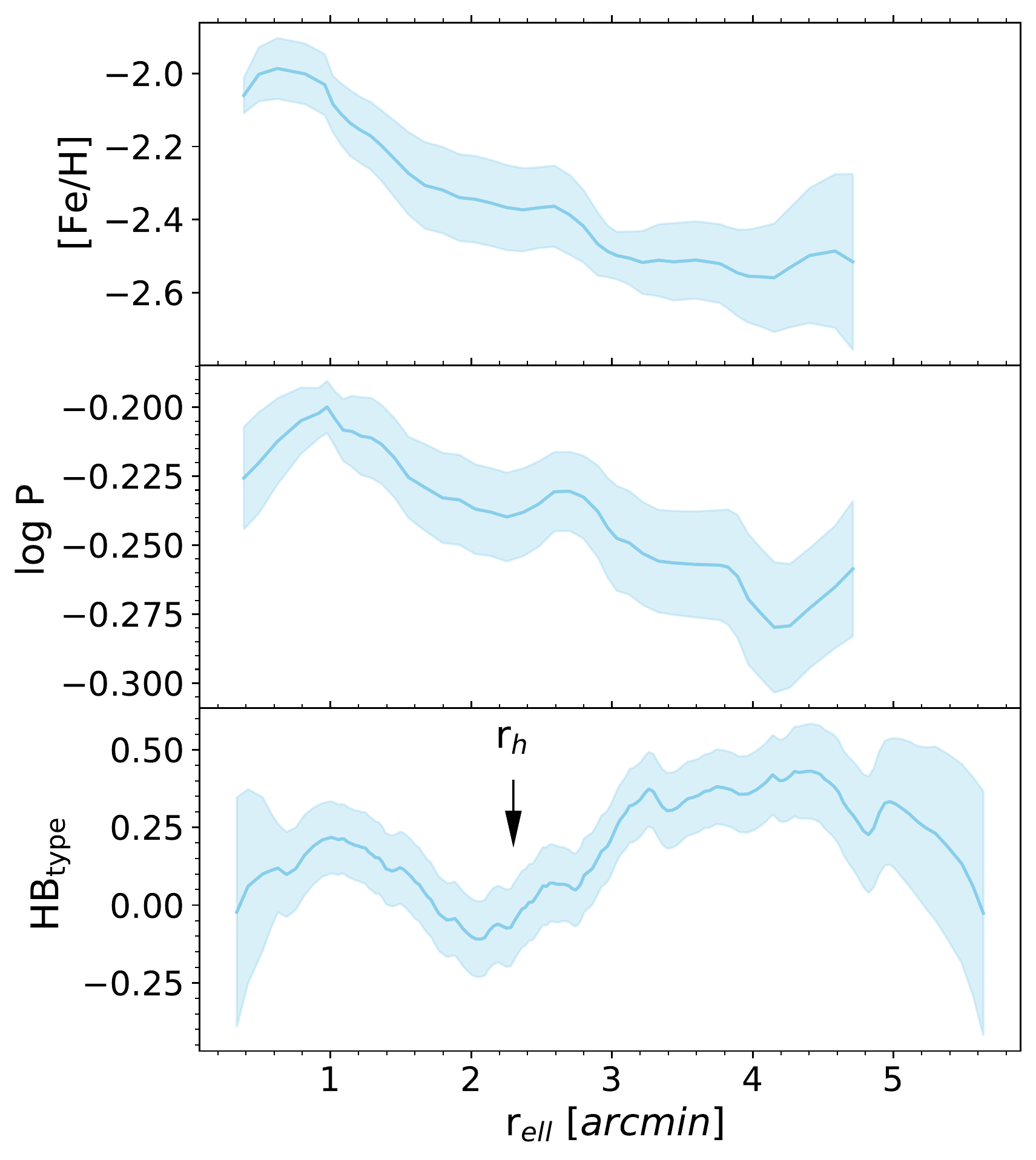} 
    \caption{Radial gradient of the metallicity (top), the logarithm of the period (middle), fundamentalized for the RRc types, of the RRL stars in Eri~II, and the morphology of the HB (HB$_{\rm type}$, bottom). The blue shaded areas show the corresponding uncertainties.}
    \label{fig:gradients}
\end{figure}

To further investigate the fast chemical enrichment of Eri~II, and taking advantage of the large portion of the galaxy covered with the DECam and Goodman data, we investigate how the properties of RRL stars change as a function of the elliptical radius (r$_{\rm ell}$) \footnote{The elliptical radius or elliptical distance, r$_{\rm ell}$, is the semi-major axis of the ellipse centred on the galaxy and defined by the morphological parameters ($\epsilon$ and PA) at the location of each individual star.}. In particular, we performed a moving average over the metallicity and the logarithm of the period (fundamentalized for RRc stars) for the clean sample of RRL stars in Eri~II. The top panel of Figure~\ref{fig:gradients} presents how the RRL metallicity changes with radius. The plot shows that the metallicity of RRL stars attain the highest values ([Fe/H] $\sim -2$) in the innermost 1\arcmin, and then steadily decreases by $\sim 0.6$ dex at a distance of 4\arcmin (or 430 pc). The inner region (0.5\arcmin to 3\arcmin) is characterized by a gradient of $-0.199 \pm 0.014$~dex/arcmin while the outermost region (3\arcmin to 4.7\arcmin) has a milder negative gradient of $-0.007 \pm 0.016$~dex/arcmin.

To check whether this gradient can be an artifact of the PL($i$) adopted to derive the metallicity, the middle and lower panels of Figure~\ref{fig:gradients} show the trend  with radius of the logarithm of the fundamentalized period and of the HB$_{\rm type}$ index\footnote{The HB$_{\rm type}$ index parametrizes the morphology of the HB and is defined as ($B$-$R$)/($B$+$V$+$R$), \citep{Lee1990}, where B and R are the number of HB stars bluer and redder than the instability strip, respectively, and V is the number of RRL stars.}, respectively. The period shows a very similar tendency as the metallicity, with a steady decrease of $\sim 0.1$~d. How to explain a trend in the period with the position within the galaxy? While a problem with the photometry or the photometric calibration could easily introduce an error in the luminosity of the stars, it is very difficult to find a systematic effect that can alter the period. In addition, this period gradient also discards the possibility that the metallicity gradient observed in Eri~II is a consequence of the inclination and geometry of the galaxy along the line of sight, which could affect the absolute magnitude of the RRL stars used to determine the metallicity.

\begin{figure}
    \hspace{-0.6cm}
    \includegraphics[width=0.50\textwidth]{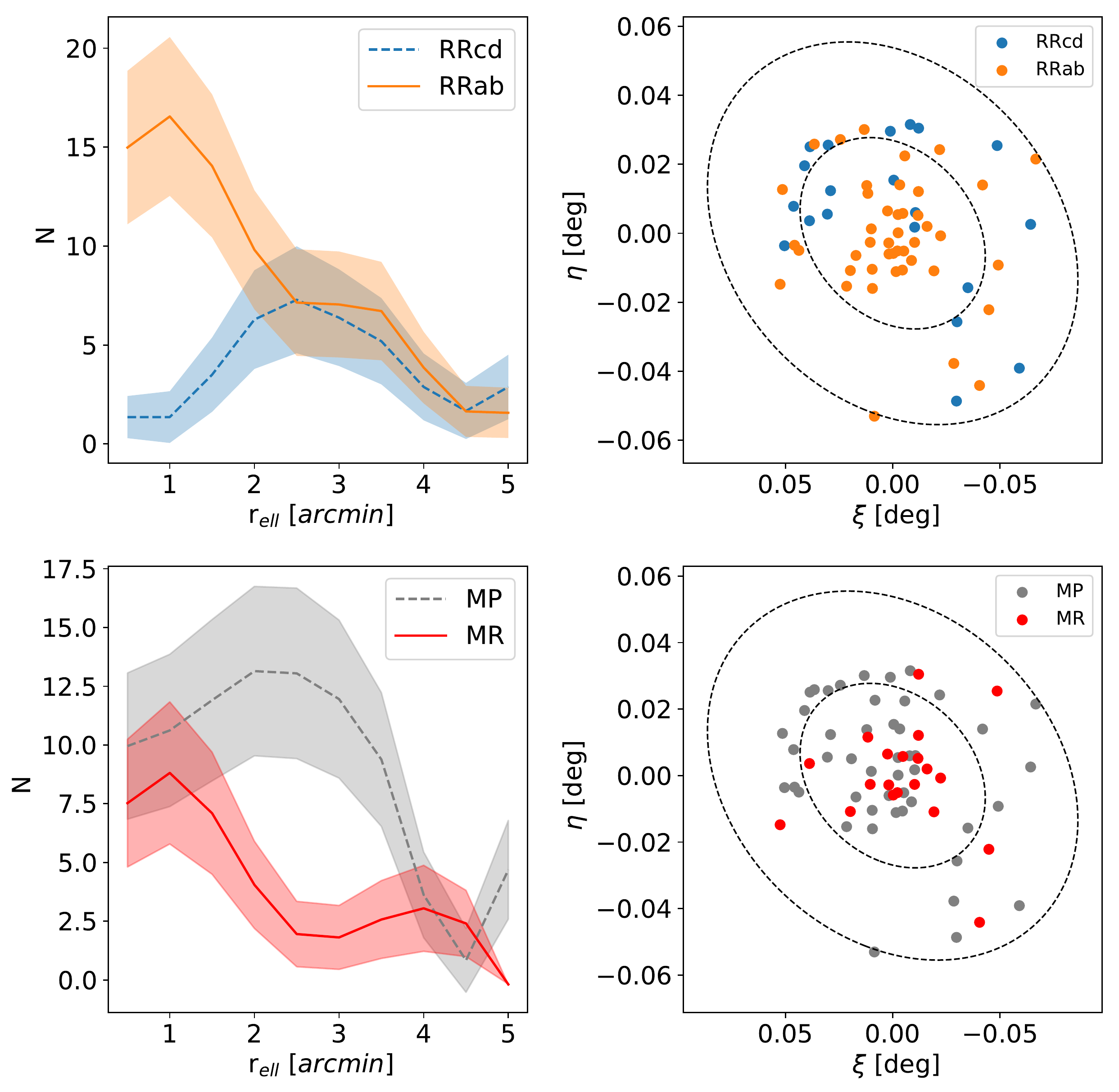} 
    \caption{\textit{Top-left}. Number of RRab and number of RRc+RRd as a function of the elliptical distance (r$_{\rm ell}$). \textit{Top-right}. Spatial distribution for the RRab and RRc+RRd. There is a clear distinction between the distribution of RRab (centrally concentrated) and the distribution of RRc+RRd (located in the outer parts of Eri~II). \textit{Bottom-left}. Number of the more metal-rich (MR, [Fe/H] $> -2.4$) and number of the more metal-poor (MP, [Fe/H] $< -2.4$) RRL stars as a function of the elliptical distance. \textit{Bottom-right}. Spatial distribution for the MR and MP RRL stars. The MR are clearly more centrally concentrated than the MP RRL stars, that are spread over the entire body of Eri~II. The ellipses in the right panels mark the locus corresponding to 1~r$_{\rm h}$ and 2~r$_{\rm h}$ of Eri~II using the morphological parameters calculated in this work.}
    \label{fig:profiles}
\end{figure}

Furthermore, the bottom panel shows that overall the HB$_{\rm type}$ increases for r$_{\rm ell}$ $>$2~r$_{\rm h}$, meaning that the HB gets bluer in the outskirts of Eri~II. This is supported by Figure~\ref{fig:profiles}, which shows in the top panels the number of RRab and RRcd \textit{vs.} radius (left) and their spatial distribution. The number of the redder RRab stars decreases with the distance, while the number of the RRcd bluer stars has its maximum between 3\arcmin and 4\arcmin from the center. The top right panel shows that indeed RRab stars cluster in the innermost regions of Eri~II, while RRc stars are dominant between 1 and 2~r$_{\rm h}$. 

The bottom panels of Figure~\ref{fig:profiles} show a similar plot but splitting RRL stars in two samples with approximately the same number of objects according to their metallicity. To do this we use eq.~\ref{eq:pli_eri2} to split the sample into a bright (metal-poor or MP) and faint (metal-rich or MR) sub-sample. By definition, the MP sample is represented by the RRL stars with a metallicity smaller than $-2.4$~dex and the MR sample are the RRL stars more metal-rich than $-2.4$~dex. The MR sample (red) is more centrally concentrated while the MP sample (grey) is more spread over the entire body of Eri~II. In addition, the MR sample closely follows the behaviour of the RRab stars, while the more MP sample mostly corresponds to the RRcd stars.

Overall then, this analysis strongly supports a  metallicity gradient for the RRL stars of Eri~II. The decreasing metallicity towards the outskirts is driving the change in the HB morphology: a more metal-poor population implies a bluer HB when moving outward, which in turn shows up in the higher fraction of RRcd stars and in the shorter mean period.

\section{Summary and final remarks}\label{sec:conclusions}

We present in this work multi-epoch $g,r,i$ SOAR/Goodman and Blanco/DECam, and $F475W, F606W, F814W$ ACS/HST photometry of Eri~II. 

Using these data, we derive new and independent structural parameters (RA, Dec, $\epsilon$, PA) for Eri~II and its star cluster. Overall, they are in very good agreement with those reported by \cite{Crnojevic2016b}. We also agree with \cite{Simon2021}, except for the r$_{\rm h}$ and the RA of the galaxy's central coordinates.

We perform the most comprehensive study of variable stars in this galaxy. We detect 69 variable stars in total, 68 in the Goodman+DECam photometry and 43 in ACS. Out of them, 67 are RRL stars and 2 are ACs. Only 7 of them (5 RRL and 2 AC stars) are detected in previous work by \citet{Stringer2020}.

The position of the two ACs is particularly interesting. One is located very close to the center while the other is the farthest variable star of this galaxy. The probability of them being foreground RRL stars is very low. The two AC stars (with masses of $\sim$1.2~M$_{\odot}$) are compatible with being F (V34) and FO (V63) pulsators based on the period-Wesenheit and the mass classification. 

From the sample of RRL stars, 44 are RRab, 12 RRc, 8 probable RRd, and 3 \textit{peculiar} RRL stars that are brighter than the bulk of the RRL stars, similar to those found in Sculptor \citep{MartinezVazquez2016b} and Carina \citep{Coppola2015}. By looking at the ratio of RRc+RRd stars ($f_{cd}$ = 0.31) and the mean periods of the RRab (0.62~d) and RRc (0.34~d) type, Eri~II can be classified as an Oosterhoff-intermediate system like the classical dwarf galaxies. However, in more detail, following \cite{MartinezVazquez2017}, we analyze the percentage of RRab close to the Oo-I and Oo-II loci and establish that 66\% of the RRab are Oo-I and 34\% are Oo-II. This ratio is also seen in other systems (such us Ursa Major~I, Sextans and Tucana, for example; see Table 6 in \citealt{MartinezVazquez2017}) and suggests that the traditional Oosterhoff classification \citep[see e.g.,][]{Catelan2009} is insufficient to account for the complexity of the early evolution of dwarf galaxies as noticed previously by other works \citep{Kuehn2013b, MartinezVazquez2016b}.

We use the information of the amplitudes of the RRL stars in the different bands to obtain the amplitude ratios, which are useful for future comparisons and transformations between SDSS-like filters, such as DECam and the forthcoming set of filters in which the LSST survey will be carried out. The relations obtained appear as eqs.~\ref{eq:AgAr}, \ref{eq:AgAi}, and \ref{eq:ArAi}. We see that these empirical ratios are in good agreement with those obtained from the SDSS theoretical RRL star models of \citet{Marconi2006}.

Additionally we derive the PL relations for the RRL stars in Eri~II. These are the first PL relations obtained for a low metallicity system in the DECam filters and could be used in systems with similar metallicity to Eri~II (i.e., in metal-poor systems with [Fe/H] $> -2.0$). In particular, our derived PL($i$) is in very good agreement with that predicted theoretically by \cite{Caceres&Catelan2008}.

We use PL relations for the RRL stars to measure the true distance modulus to Eri~II. The results of the different relations are in a $>$0.5$\sigma$ agreement. Therefore, we adopted as the true distance modulus, the average of those distance moduli, i.e., $\mu_0 = 22.84 \pm 0.05~(\sigma_{\rm rand}= 0.01)$~mag (D$_{\odot} = 370 \pm 9$~kpc). 

We obtain the metallicity distribution of the RRL stars in Eri~II, which is in agreement with the metallicities obtained by \citealt{Li2017} based on 16 RGB stars. The overall low metallicity of Eri~II is also consistent with the RGB bump location, 0.65 mag brighter than the HB level. The dispersion of the metallicity distribution derived from the RRL stars ($\sigma_{\rm [Fe/H], RRL} = 0.3$~dex, 0.2~dex intrinsic) is slightly smaller than the derived from the spectroscopic studies \citep{Li2017, Zoutendijk2020}. Nevertheless, this could be at least partly explained by assuming that the spectroscopic studies could include RGB stars slightly younger than RRL stars -- not excluded by the SFH determinations \citep[][]{Simon2021, Gallart2021}, with no counterpart in the RRL stars -- and also low metallicity stars not producing RRL stars. Overall, the value of the metallicity dispersion obtained from the RRL stars in Eri~II is a very interesting result, since it is a further confirmation that this low mass galaxy was substantially enriched in its very short star formation burst \citep[likely shorter than 1~Gyr,][]{Gallart2021}, similar to what also occurred in much more massive galaxies such as Sculptor dSph \citep[][]{MartinezVazquez2016a}. 

We perform a HB simulation by selecting two $\alpha$-enhanced BaSTI stellar models corresponding to [Fe/H]$= -2.62$ and [Fe/H]$= -2.14$, together with the RRL metallicity distribution of this work, the SFH derived by \citet{Gallart2021}, and considering the same number of synthetic HB stars as the observed HB stars of Eri~II. The synthetic HB shows a fair agreement with the observed HB of Eri~II and in particular shows that a metallicity spread is needed to replicate the colour range covered by the Eri~II HB stars. In addition, this HB simulation reproduces the location of the two main peaks in the observed RRL period distribution.

Finally, the study of the metallicity gradients and the spatial distribution of RRL stars yields interesting findings regarding the stellar population gradients of this small and old system. We find that the RRab stars are more concentrated than the RRc stars, mainly located outside $\sim 1$~r$_{\rm h}$. Moreover, metal-poor RRL stars are spread around Eri~II while the metal-rich stars are centrally concentrated. This behavior is similar to what is seen in galaxies like Sculptor \citep[][see their Figure 4]{MartinezVazquez2016a} and it is in agreement with outside-in galaxy formation scenarios.

As shown in this paper, Eri~II is a very interesting galaxy in terms of its fast and early chemical enrichment. It is surprising how such a low-mass galaxy can harbour a strong metallicity gradient ($-0.199 \pm 0.014$~dex/arcmin between 0.5\arcmin and 3\arcmin) at its early epochs, comparable with galaxies like Sculptor (which is $\sim 50$ times more massive). This makes it mandatory to explore the early metal-enrichment of low-mass dwarf galaxies in more detail. Being able to answer questions relating to the formation sites for these galaxies, and ascertaining the main mechanisms that drove their fast and early evolution are essential to understand the nature, formation and evolution of these systems.  Clearly, follow-up studies are needed to understand more deeply the big picture of these galaxies as part of the MW and LG zoo. 

\section*{Acknowledgements}

We thank our anonymous referee for an extensive report that helped improve this paper.
This research has been supported by the Spanish Ministry of Economy and Competitiveness (MINECO) under the grant AYA2017-89076-P. SC acknowledges support from Premiale INAF MITiC, from Istituto Nazionale di Fisica Nucleare (INFN) (Iniziativa specifica TAsP), and project INAF Mainstream (PI. S. Cassisi). SS acknowledges support from the ERC Starting Grant NEFERTITI H2020/808240 and the PRIN-MIUR2017, The quest for the first stars, prot. n. 2017T4ARJ5. EB acknowledges financial support from a Vici grant from the Netherlands Organisation for Scientific Research (NWO).

Based on observations obtained at the Southern Astrophysical Research (SOAR) telescope (NOIRLab Prop. ID 2018A-0310; PI: C.~E.~Mart\'inez-V\'azquez), which is a joint project of the Minist\'{e}rio da Ci\^{e}ncia, Tecnologia e Inova\c{c}\~{o}es (MCTI/LNA) do Brasil, the US National Science Foundation's NOIRLab, the University of North Carolina at Chapel Hill (UNC), and Michigan State University (MSU).

Based on observations at Cerro Tololo Inter-American Observatory, NSF's NOIRLab (NOIRLab Prop. ID 2018A-0310; PI: C.~E.~Mart\'inez-V\'azquez), which is managed by the Association of Universities for Research in Astronomy (AURA) under a cooperative agreement with the National Science Foundation.

This project used data obtained with the Dark Energy Camera (DECam), which was constructed by the Dark Energy Survey (DES) collaboration. Funding for the DES Projects has been provided by the US Department of Energy, the US National Science Foundation, the Ministry of Science and Education of Spain, the Science and Technology Facilities Council of the United Kingdom, the Higher Education Funding Council for England, the National Center for Supercomputing Applications at the University of Illinois at Urbana-Champaign, the Kavli Institute for Cosmological Physics at the University of Chicago, Center for Cosmology and Astro-Particle Physics at the Ohio State University, the Mitchell Institute for Fundamental Physics and Astronomy at Texas A\&M University, Financiadora de Estudos e Projetos, Funda\c{c}\~{a}o Carlos Chagas Filho de Amparo \`a Pesquisa do Estado do Rio de Janeiro, Conselho Nacional de Desenvolvimento Cient\'ifico e Tecnol\'ogico and the Minist\'erio da Ci\^encia, Tecnologia e Inova\c{c}\~{a}o, the Deutsche Forschungsgemeinschaft and the Collaborating Institutions in the Dark Energy Survey.

The Collaborating Institutions are Argonne National Laboratory, the University of California at Santa Cruz, the University of Cambridge, Centro de Investigaciones Energ\'eticas, Medioambientales y Tecnol\'ogicas-Madrid, the University of Chicago, University College London, the DES-Brazil Consortium, the University of Edinburgh, the Eidgen\"ossische Technische Hochschule (ETH) Z\"urich, Fermi National Accelerator Laboratory, the University of Illinois at Urbana-Champaign, the Institut de Ci\`encies de l'Espai (IEEC/CSIC), the Institut de F\'isica d'Altes Energies, Lawrence Berkeley National Laboratory, the Ludwig-Maximilians Universit\"at M\"unchen and the associated Excellence Cluster Universe, the University of Michigan, NSF's NOIRLab, the University of Nottingham, the Ohio State University, the OzDES Membership Consortium, the University of Pennsylvania, the University of Portsmouth, SLAC National Accelerator Laboratory, Stanford University, the University of Sussex, and Texas A\&M University.

\section*{Data Availability}

The data underlying this article are available in the article and in its online supplementary material. 

Tables~\ref{tab:puls_prop}, \ref{tab:comments}, and the full version of Table~\ref{tab:photometry} are published in electronic form.

The full set of light curves in the ACS and DECam filters for the 41 RRL stars in ACS and the 66 RRL stars in Goodman+DECam is available as Support Information Material in the online version of the journal.



\bibliographystyle{mnras}

\newcommand{\noop}[1]{}



\appendix
\section{Pulsation parameters and classification for the variable stars detected in Eri~II}\label{ap:puls_param}

Table~\ref{tab:puls_prop} gives the positions, periods, intensity-weighted magnitudes and amplitudes for the filters $i, r, g, F814W, F606W, F475W$ and classifications for the 69 variable stars detected in Eri~II.

\input{vartab_final.tex}

\section{Comments on individual variable stars detected in Eri~II}\label{ap:comments}

Table~\ref{tab:comments} lists the notes we have made on some individual variable stars based on the visualization of their light curves.

\input{comments_vartab.tex}

\section{Time-series photometry of the variable stars in Eri~II}\label{ap:photometry}

Table~\ref{tab:photometry} shows the individual $i, r, g, F814W, F606W, F475W$ photometry for the variables detected in this work. Table~\ref{tab:photometry} is published in its entirety as supplementary material (online) in the journal. A portion is shown here for guidance regarding its form and content.

\input{photometry_variables_EriII_tab_red.tex}


\bsp	
\label{lastpage}
\end{document}

%% file: morph_param_tab.tex
\begin{table*}
\small
\caption{Morphological parameters.}
\label{tab:morph_param}
\begin{tabular}{lccccccc}
\hline
Sample & RA & DEC & $\epsilon$ & PA     &   r$_h$   & $\mu_0$ & D$_{\sun}^{(e)}$  \\
       & (deg) & (deg) &       & (deg)  & (arcmin)  & (mag)  &  (kpc) \\
\hline
\multicolumn{8}{l}{\textbf{Eridanus~II}} \\
\hline
\citet{Crnojevic2016b} & 56.084 $\pm$ 0.003 & $-$43.534 $\pm$ 0.001 & 0.48 $\pm$ 0.04 & 72.6 $\pm$ 3.3 & 2.31 $\pm$ 0.12 & 22.8 $\pm$ 0.1$^{(c)}$ & 363 $\pm$ 17 \\
\citet{Simon2021} &  56.0880 $\pm$ 0.0005 & $-$43.5334 $\pm$ 0.0002 & 0.45 $\pm$ 0.02 & 77.8 $\pm$ 1.2 & 3.03 $\pm$ 0.12$^{(a)}$ & 22.65 $\pm$ 0.08$^{(c)}$ & 339 $\pm$ 12 \\
This work & 56.0829 $\pm$ 0.0015 & $-$43.535 $\pm$ 0.001 & 0.40 $\pm$ 0.03 & 75 $\pm$ 3 & 2.30 $\pm$ 0.12 & 22.84 $\pm$ 0.07$^{(d)}$ & 370 $\pm$ 12 \\
RR Lyrae stars$^{(b)}$ & 56.081 $\pm$ 0.001 & $-$43.533 $\pm$ 0.001 & 0.50 $\pm$ 0.08 & 80 $\pm$ 4 & ... \\
\hline
\multicolumn{8}{l}{\textbf{Cluster}} \\
\hline 
\citet{Crnojevic2016b} & 56.0925 $\pm$ 0.0003 & $-$43.5331 $\pm$ 0.0006 & ... &  ...  & 0.11 $\pm$ 0.01 & ... & ...  \\
\citet{Simon2021} & 56.0933 $\pm$ 0.0001 & $-$43.5334 $\pm$ 0.0001 & 0.31 $\substack{+0.05\\-0.06}$ & 75 $\pm$ 6 & 0.16 $\pm$ 0.01 & ... & ... \\
This work & 56.0933 $\pm$ 0.0002 & $-$43.5334 $\pm$ 0.0001 &  0.40 $\pm$ 0.06 & 74 $\pm$ 6 & 0.20 $\pm$ 0.02 &  & \\
\hline
\hline
\end{tabular}
\begin{tablenotes}
\item $^{(a)}$ \citet{Simon2021} acknowledge that the smaller size given by \citet{Crnojevic2016b} is more accurate. 
\item $^{(b)}$ Parameters based on a bivariate Gaussian distribution of the RR Lyrae variable stars using the routine \verb+fit_bivariate_normal+ in AstroML \citep{Ivezic2014}.
\item $^{(c)}$ Based on isochrone-fitting.
\item $^{(d)}$ Based on the RR Lyrae stars (see \S~\ref{sec:distance})
\item $^{(e)}$ Heliocentric distance converted from the distance modulus ($\mu_0$) in column 7.
\end{tablenotes}
\end{table*} 

%% file: mass_acep_tab.tex
\begin{table}
\begin{scriptsize}
\centering
\caption{Parameters of the AC stars in Eri~II.} 
\label{tab:mass_acep}
\begin{tabular}{lccccc} 
\hline
ID & M$_{\rm{MPLA, FU}}$ & M$_{\rm{MPLC, FU}}$ & M$_{\rm{MPLC, FO}}$ & MODE$_{\rm{mass}}$ & M$^{(a)}$ \\
   & (M$_{\sun}$) & (M$_{\sun}$) & (M$_{\sun}$) & & (M$_{\sun}$) \\ 
\hline 
V34 & 1.22 $\pm$ 0.14 & 1.23 $\pm$ 0.06 & 0.72 $\pm$ 0.03 & F & 1.23 $\pm$ 0.06 \\
V63 & 1.60 $\pm$ 0.19 & 1.91 $\pm$ 0.09 & 1.17 $\pm$ 0.05 & FO & 1.17 $\pm$ 0.05 \\
\hline
\hline
\end{tabular}
\begin{tablenotes}
\item $^{(a)}$ Final value of the mass. The mass of the fundamental mode is the average value of the M$_{\rm{MPLA, FU}}$ and M$_{\rm{MPLC, FU}}$ measurements.
\end{tablenotes}
\end{scriptsize}
\end{table}

%% file: rrl_summary_tab.tex
\begin{table*}
\centering
\caption{Mean properties of the RR Lyrae stars in Eri~II.}
\label{tab:summary_rrl}
\resizebox{1.0\textwidth}{!}{
\begin{tabular}{lrcccccccccccccc}
\hline
Pulsator   & N  & $\langle$P$\rangle$ & $\sigma_{\langle P \rangle}$ & $\langle$ i $\rangle$ & $\sigma_{\langle i \rangle}$ & $\langle$ r $\rangle$ & $\sigma_{\langle r \rangle}$ & $\langle$ g $\rangle$ & $\sigma_{\langle g \rangle}$ & $\langle$ F814W $\rangle$ & $\sigma_{\langle F814W \rangle}$ & $\langle$ F606W $\rangle$ & $\sigma_{\langle F606W \rangle}$ & $\langle$ F475W $\rangle$ & $\sigma_{\langle F475W \rangle}$\\   
\hline
RRab       & 44 & 0.62  & 0.01 & 23.14 & 0.01 & 23.14 & 0.01 & 23.31 & 0.01 & 22.71 & 0.01 & 23.11 & 0.01 & 23.47 & 0.02  \\
RRc        & 12 & 0.34  & 0.01 & 23.25 & 0.03 & 23.20 & 0.02 & 23.25 & 0.03 & 22.84 & 0.02 & 23.11 & 0.04 & 23.39 & 0.03  \\
RRd        &  8 & 0.38  & 0.01 & 23.21 & 0.02 & 23.20 & 0.02 & 23.31 & 0.03 & 22.77 & 0.02 & 23.09 & 0.02 & 23.45 & 0.03  \\
Peculiar   &  3 & 0.58  & 0.08 & 22.82 & 0.03 & 22.83 & 0.04 & 23.01 & 0.11 & 22.48 & 0.11 & 22.87 & 0.13 & 23.18 & 0.17  \\
\hline
\hline
\end{tabular}
}
\end{table*}

%% file: distance_moduli_tab.tex
\begin{table}
\centering
\small
\caption{Distance moduli to Eri~II from its RRL stars.}
\label{tab:distance}
\begin{tabular}{lccccc}
\hline
PL relation & N & $\mu_0$ & $\sigma_{\rm sys}$ & $\sigma_{\rm rand}$ & D$_\odot$ \\
            &   & (mag)   & (mag)              & (mag)               & (kpc)     \\ 
\hline
$i$-SDSS    & 46 & 22.83  & 0.10               &  0.01               & 368 $\pm$ 17 \\
$F814$      & 29 & 22.83  & 0.09               &  0.01               & 368 $\pm$ 15 \\
$I$         & 46 & 22.85  & 0.08               &  0.01               & 372 $\pm$ 14 \\
\hline
            & & $\langle \mu_0 \rangle $ & $\sigma_{\langle \mu_0 \rangle}$ & $\sigma_{\rm rand}$ & $\langle $D$_\odot \rangle$ \\
            & & (mag)   & (mag)              & (mag)    & (kpc) \\ 
\hline
Average     & &  22.84  & 0.05               &  0.01  & 370 $\pm$ 9 \\
\hline
\hline
\end{tabular}
\end{table} 

%% file: vartab_final.tex
\begin{table*}
\caption{Pulsation properties of the variables in Eri~II}
\label{tab:puls_prop}
\resizebox{1.0\textwidth}{!}{
\begin{tabular}{|l|r|r|r|r|r|r|r|r|r|r|r|r|r|r|r|l}
\hline
  \multicolumn{1}{|c|}{Name} &
  \multicolumn{1}{c|}{RA} &
  \multicolumn{1}{c|}{Dec} &
  \multicolumn{1}{c|}{Period} &
  \multicolumn{1}{c|}{$\langle$i$\rangle$} &
  \multicolumn{1}{c|}{$\Delta_i$} &
  \multicolumn{1}{c|}{$\langle$r$\rangle$} &
  \multicolumn{1}{c|}{$\Delta_r$} &
  \multicolumn{1}{c|}{$\langle$g$\rangle$} &
  \multicolumn{1}{c|}{$\Delta_g$} &
  \multicolumn{1}{c|}{$\langle$F814W$\rangle$} &
  \multicolumn{1}{c|}{$\Delta_{F814W}$} &
  \multicolumn{1}{c|}{$\langle$F606W$\rangle$} &
  \multicolumn{1}{c|}{$\Delta_{F606W}$} &
  \multicolumn{1}{c|}{$\langle$F475W$\rangle$} &
  \multicolumn{1}{c|}{$\Delta_{F475W}$} &
  \multicolumn{1}{l|}{Type} \\
  \multicolumn{1}{|c|}{ } &
  \multicolumn{1}{c|}{(deg)} &
  \multicolumn{1}{c|}{(deg)} &
  \multicolumn{1}{c|}{(d)} &
  \multicolumn{1}{c|}{(mag)} &
  \multicolumn{1}{c|}{(mag)} &
  \multicolumn{1}{c|}{(mag)} &
  \multicolumn{1}{c|}{(mag)} &
  \multicolumn{1}{c|}{(mag)} &
  \multicolumn{1}{c|}{(mag)} &
  \multicolumn{1}{c|}{(mag)} &
  \multicolumn{1}{c|}{(mag)} &
  \multicolumn{1}{c|}{(mag)} &
  \multicolumn{1}{c|}{(mag)} &
  \multicolumn{1}{c|}{(mag)} &
  \multicolumn{1}{c|}{(mag)} &
  \multicolumn{1}{c|}{} \\
 \hline
V01  &  55.990914  & $-$43.513449  &  0.6963051  & 23.00 & 0.43 & 23.03 & 0.56 & 23.23 & 0.90 &		  &		 &		  &		 &		  &		 &  RRab  \\
V02  &  55.994195  & $-$43.532382  &  0.3671909  & 23.03 & 0.45 & 23.00 & 0.66 & 23.04 & 0.94 &		  &		 &		  &		 &		  &		 &  RRc   \\
V03  &  56.001382  & $-$43.574081  &  0.4423017  & 23.12 & 0.23 & 23.09 & 0.28 & 23.22 & 0.39 &		  &		 &		  &		 &		  &		 &  RRd   \\
V04  &  56.015008  & $-$43.544199  &  0.6217043  & 23.05 & 0.69 & 23.05 & 1.02 & 23.16 & 1.36 &		  &		 &		  &		 &		  &		 &  RRab  \\
V05  &  56.015761  & $-$43.509555  &  0.3658020  & 23.26 & 0.38 & 23.22 & 0.45 & 23.33 & 0.57 &		  &		 &		  &		 &		  &		 &  RRc   \\
V06  &  56.021015  & $-$43.557150  &  0.6495050  & 23.13 & 0.43 & 23.14 & 0.56 & 23.34 & 0.84 &		  &		 &		  &		 &		  &		 &  RRab  \\
V07  &  56.025117  & $-$43.520985  &  0.5743040  & 23.12 & 0.59 & 23.12 & 0.71 & 23.31 & 1.07 &		  &		 &		  &		 &		  &		 &  RRab  \\
V08  &  56.026984  & $-$43.579114  &  0.5394511  & 23.23 & 0.66 & 23.23 & 1.05 & 23.35 & 1.23 &		  &		 &		  &		 &		  &		 &  RRab  \\
V09  &  56.034517  & $-$43.550770  &  0.3819020  & 23.21 & 0.22 & 23.22 & 0.37 & 23.45 & 0.53 &		  &		 &		  &		 &		  &		 &  RRd   \\
V10  &  56.041507  & $-$43.560660  &  0.2769208  & 23.30 & 0.23 & 23.25 & 0.33 & 23.40 & 0.54 &		  &		 &		  &		 &		  &		 &  RRc   \\
V11  &  56.041829  & $-$43.583645  &  0.2835108  & 23.30 & 0.16 & 23.20 & 0.28 & 23.21 & 0.29 &		  &		 &		  &		 &		  &		 &  RRc   \\
V12  &  56.043558  & $-$43.572738  &  0.6595050  & 23.05 & 0.52 & 23.04 & 0.64 & 23.17 & 1.09 &		  &		 &		  &		 &		  &		 &  RRab  \\
V13  &  56.052077  & $-$43.535731  &  0.5389422  & 23.29 & 0.85 & 23.28 & 1.10 & 23.44 & 1.21 &	22.92 &	0.62 &	23.29 &	1.01 &	23.55 &	0.76 &  RRab  \\
V14  &  56.052716  & $-$43.510743  &  0.5701511  & 23.16 & 0.60 & 23.14 & 0.88 & 23.24 & 1.26 &	22.74 &	0.59 &	23.10 &	0.99 &		  &		 &  RRab  \\
V15  &  56.056346  & $-$43.545898  &  0.5928861  & 23.16 & 0.43 & 23.16 & 0.52 & 23.33 & 0.77 &	22.68 &	0.53 &		  &		 &	23.38 &	1.32 &  RRab  \\
V16  &  56.060786  & $-$43.533004  &  0.5741372  & 23.20 & 0.55 & 23.19 & 0.62 & 23.36 & 0.99 &	22.73 &	0.53 &	23.14 &	0.74 &	23.42 &	0.91 &  RRab  \\
V17  &  56.066222  & $-$43.504528  &  0.2979708  & 23.34 & 0.35 & 23.28 & 0.41 & 23.26 & 0.59 &		  &		 &		  &		 &		  &		 &  RRc   \\
V18  &  56.066335  & $-$43.522899  &  0.6167251  & 23.16 & 0.39 & 23.18 & 0.50 & 23.37 & 0.68 &	22.69 &	0.41 &	23.10 &	0.57 & $<$23.67 &	 &  RRab  \\
V19  &  56.066611  & $-$43.529831  &  0.6097011  & 23.21 & 0.48 & 23.22 & 0.57 & 23.43 & 0.72 &	22.75 &	0.54 &	23.17 &	0.62 &	23.66 &	0.66 &  RRab  \\
V20  &  56.068276  & $-$43.528991  &  0.3872109  & 23.34 & 0.17 & 23.31 & 0.21 & 23.44 & 0.33 &	22.82 &	0.24 &	23.09 &	0.50 &	23.53 &	0.39 &  RRd   \\
V21  &  56.068753  & $-$43.533239  &  0.3826109  & 23.25 & 0.23 & 23.21 & 0.34 & 23.33 & 0.51 &	22.79 &	0.28 &	23.15 &	0.36 &	23.43 &	0.48 &  RRd   \\
V22  &  56.068808  & $-$43.537646  &  0.6156011  & 23.20 & 0.45 & 23.17 & 0.56 & 23.37 & 0.86 &	22.76 &	0.37 &	23.18 &	0.51 & $<$23.69 &	 &  RRab* \\
V23  &  56.070792  & $-$43.542879  &  0.7555413  &       &      &       &      &       &      & 22.56 & 0.26 &  23.05 & 0.46 &  23.40 & 0.48 &  RRab  \\
V24  &  56.071604  & $-$43.503462  &  0.3776435  & 23.19 & 0.35 & 23.21 & 0.39 & 23.31 & 0.54 &		  &		 &		  &		 &		  &		 &  RRd   \\
V25  &  56.072021  & $-$43.529039  &  0.6196111  & 22.87 & 0.36 & 22.90 & 0.50 & 23.21 & 0.68 &	22.70 &	0.31 &	23.12 &	0.60 &	23.48 &	0.71 &  \textit{peculiar} RRL  \\
V26  &  56.075107  & $-$43.512568  &  0.6271212  & 22.96 & 0.35 & 22.99 & 0.46 & 23.22 & 0.65 &	22.72 &	0.43 &	23.13 &	0.59 &	23.52 &	0.75 &  RRab  \\
V27  &  56.075765  & $-$43.540154  &  0.6393111  & 23.19 & 0.35 & 23.20 & 0.45 & 23.42 & 0.66 &	22.72 &	0.32 &	23.07 &	0.50 &	23.56 &	0.69 &  RRab  \\
V28  &  56.076349  & $-$43.529244  &  0.6035311  & 23.20 & 0.59 & 23.18 & 0.81 & 23.34 & 1.26 &	22.73 &	0.56 &	23.10 &	0.92 &	23.45 &	1.30 &  RRab  \\
V29  &  56.076600  & $-$43.545661  &  0.7215613  & 23.06 & 0.41 & 23.07 & 0.61 & 23.27 & 0.87 &	22.59 &	0.44 &	22.99 &	0.76 &	23.31 &	0.75 &  RRab  \\
V30  &  56.078351  & $-$43.520961  &  0.6474112  & 23.02 & 0.29 & 23.08 & 0.45 & 23.26 & 0.58 &	22.69 &	0.35 &	23.09 &	0.52 & $<$23.70 &	 &  RRab* \\
V31  &  56.079398  & $-$43.534866  &  0.6515512  & 23.11 & 0.34 & 23.11 & 0.47 & 23.30 & 0.76 &	22.62 &	0.46 &	23.05 &	0.62 &	23.41 &	1.08 &  RRab  \\
V32  &  56.079464  & $-$43.529579  &  0.7099512  & 23.07 & 0.28 & 23.10 & 0.35 & 23.31 & 0.50 &	22.62 &	0.29 &	23.04 &	0.40 & $<$23.53 &	 &  RRab  \\
V33  &  56.079877  & $-$43.540129  &  0.5324321  & 23.27 & 0.66 & 23.22 & 0.99 & 23.32 & 1.36 &	22.87 &	0.79 &	23.19 &	1.22 &		  &		 &  RRab  \\
V34  &  56.080364  & $-$43.544475  &  0.8887514  & 22.21 & 0.55 & 22.24 & 0.73 & 22.37 & 1.00 &	21.74 &	0.60 &	22.13 &	0.73 &	22.45 &	0.96 &  AC	  \\
V35  &  56.080736  & $-$43.546118  &  0.6084580  & 23.15 & 0.42 & 23.14 & 0.60 & 23.31 & 0.81 &	22.72 &	0.54 &	23.12 &	0.63 &	23.47 &	0.79 &  RRab  \\
V36  &  56.082197  & $-$43.519593  &  0.3403509  & 23.20 & 0.30 & 23.12 & 0.42 & 23.13 & 0.61 &	22.73 &	0.28 &	22.96 &	0.40 &	23.27 &	0.51 &  RRc   \\
V37  &  56.082526  & $-$43.540848  &  0.5742922  & 23.23 & 0.61 & 23.22 & 0.83 & 23.37 & 1.25 &	22.77 &	0.66 &	23.15 &	0.83 &	23.39 &	1.14 &  RRab* \\
V38  &  56.084405  & $-$43.505423  &  0.3047611  & 23.39 & 0.30 & 23.29 & 0.53 & 23.27 & 0.67 &	22.91 &	0.39 &	23.16 &	0.50 &	23.38 &	0.56 &  RRc   \\
V39  &  56.085268  & $-$43.541011  &  0.5967652  & 23.12 & 0.48 & 23.18 & 0.59 & 23.34 & 0.82 &	22.75 &	0.46 &	23.12 &	0.85 &	23.52 &	0.72 &  RRab* \\
V40  &  56.085422  & $-$43.537809  &  0.6250511  & 23.16 & 0.55 & 23.18 & 0.55 & 23.40 & 0.85 &	22.78 &	0.30 &	23.21 &	0.43 &	23.45 &	0.94 &  RRab* \\
V41  &  56.086194  & $-$43.528525  &  0.5860652  & 23.27 & 0.55 & 23.26 & 0.66 & 23.40 & 0.91 &	22.77 &	0.62 &	23.11 &	0.88 &	23.31 &	1.20 &  RRab  \\
V42  &  56.094252  & $-$43.512333  &  0.6879912  & 22.76 & 0.31 & 22.82 & 0.46 & 23.00 & 0.71 &	22.37 &	0.31 &	22.79 &	0.49 &	23.17 &	0.77 &  \textit{peculiar} RRL  \\
V43  &  56.094682  & $-$43.588036  &  0.5553040  & 23.16 & 0.50 & 23.10 & 0.72 & 23.21 & 1.30 &		  &		 &		  &		 &		  &		 &  RRab  \\
V44  &  56.095911  & $-$43.550993  &  0.7466713  & 23.06 & 0.25 & 23.06 & 0.36 & 23.31 & 0.57 &	22.60 &	0.29 &	23.07 &	0.38 &	23.40 &	0.57 &  RRab  \\
V45  &  56.096072  & $-$43.545432  &  0.5755451  & 23.20 & 0.59 & 23.19 & 0.65 & 23.37 & 0.97 &	22.78 &	0.48 &	23.21 &	0.66 &	23.47 &	1.03 &  RRab  \\
V46  &  56.096637  & $-$43.533720  &  0.6311111  & 22.99 & 0.40 & 22.98 & 0.46 & 23.25 & 0.71 &	22.68 &	0.46 &	23.13 &	0.61 &	23.48 &	0.79 &  RRab  \\
V47  &  56.097344  & $-$43.537631  &  0.6312112  & 23.28 & 0.74 & 23.23 & 0.80 & 23.36 & 1.16 &	22.68 &	0.72 &	23.09 &	0.94 &	23.40 &	1.04 &  RRab  \\
V48  &  56.098849  & $-$43.523436  &  0.5884211  & 23.18 & 0.42 & 23.17 & 0.66 & 23.35 & 0.88 &	22.74 &	0.58 &	23.12 &	0.69 &		  &		 &  RRab  \\
V49  &  56.099573  & $-$43.521162  &  0.6935571  & 23.06 & 0.36 & 23.11 & 0.57 & 23.29 & 0.82 &	22.59 &	0.35 &	22.99 &	0.68 &	23.37 &	0.80 &  RRab  \\
V50  &  56.101174  & $-$43.504920  &  0.6353582  & 23.10 & 0.72 & 23.08 & 0.89 & 23.22 & 1.21 &	22.66 &	0.66 &	23.03 &	0.92 &		  &		 &  RRab  \\
V51  &  56.106513  & $-$43.541426  &  0.6168911  & 23.14 & 0.48 & 23.14 & 0.56 & 23.35 & 0.77 &	22.70 &	0.50 &	23.12 &	0.94 &	23.49 &	0.75 &  RRab  \\
V52  &  56.109388  & $-$43.529929  &  0.4344510  & 22.82 & 0.51 & 22.77 & 0.64 & 22.82 & 0.86 &	22.38 &	0.48 &	22.70 &	0.79 &	22.89 &	0.82 &  \textit{peculiar} RRL \\
V53  &  56.110109  & $-$43.545800  &  0.5693882  & 23.33 & 0.70 & 23.27 & 0.81 & 23.35 & 1.21 &	22.78 &	0.51 &	23.14 &	0.76 &	23.51 &	1.04 &  RRab* \\
V54  &  56.112522  & $-$43.550346  &  0.6085752  & 23.16 & 0.38 & 23.19 & 0.58 & 23.38 & 0.77 &	22.72 &	0.42 &	23.15 &	0.70 &  23.22 &	1.04 &  RRab  \\
V55  &  56.116530  & $-$43.507817  &  0.6224611  & 23.03 & 0.56 & 23.08 & 0.73 & 23.19 & 0.98 &	22.61 &	0.59 &	23.00 &	0.62 &		  &		 &  RRab  \\
V56  &  56.122841  & $-$43.522646  &  0.3871461  & 23.26 & 0.14 & 23.26 & 0.31 & 23.34 & 0.48 &	22.84 &	$\sim$0.1 &	23.21 & $\sim$0.3& $<$23.50 &	 &  RRc   \\
V57  &  56.124364  & $-$43.509429  &  0.3032100  & 23.24 & 0.16 & 23.18 & 0.23 & 23.19 & 0.32 &		  &		 &		  &		 &		  &		 &  RRd   \\
V58  &  56.124897  & $-$43.529450  &  0.4178506  & 23.16 & 0.14 & 23.18 & 0.26 & 23.25 & 0.45 &	22.71 &	0.17 &	23.03 &	$\sim$0.4 &	23.38 &	     &  RRd   \\
V59  &  56.133247  & $-$43.509141  &  0.5950040  & 23.13 & 0.42 & 23.19 & 0.64 & 23.36 & 0.88 &		  &		 &		  &		 &		  &		 &  RRab  \\
V60  &  56.136034  & $-$43.509897  &  0.3943509  & 23.16 & 0.29 & 23.16 & 0.34 & 23.27 & 0.41 &		  &		 &		  &		 &		  &		 &  RRc   \\
V61  &  56.136381  & $-$43.531359  &  0.3406042  & 23.30 & 0.42 & 23.26 & 0.53 & 23.30 & 0.73 &	22.88 &	0.40 &		  &		 &	23.43 &	0.64 &  RRc   \\
V62  &  56.139549  & $-$43.515409  &  0.3870019  & 23.16 & 0.27 & 23.17 & 0.33 & 23.27 & 0.47 &		  &		 &		  &		 &		  &		 &  RRd   \\
V63  &  56.142925  & $-$43.475216  &  0.7681060  & 22.25 & 0.32 & 22.33 & 0.40 & 22.59 & 0.64 &		  &		 &		  &		 &		  &		 &  AC	  \\
V64  &  56.143208  & $-$43.539946  &  0.5840039  & 23.14 & 0.58 & 23.13 & 0.73 & 23.30 & 1.12 &		  &		 &		  &		 &		  &		 &  RRab  \\
V65  &  56.146000  & $-$43.538458  &  0.6442046  & 23.10 & 0.49 & 23.13 & 0.58 & 23.30 & 0.89 &		  &		 &		  &		 &		  &		 &  RRab  \\
V66  &  56.146613  & $-$43.527165  &  0.3569450  & 23.21 & 0.33 & 23.20 & 0.45 & 23.28 & 0.62 &		  &		 &		  &		 &		  &		 &  RRc   \\
V67  &  56.152468  & $-$43.538592  &  0.3100060  & 23.27 & 0.31 & 23.18 & 0.31 & 23.21 & 0.51 &		  &		 &		  &		 &		  &		 &  RRc   \\
V68  &  56.153762  & $-$43.522289  &  0.7197050  & 22.94 & 0.40 & 23.00 & 0.46 & 23.23 & 0.69 &		  &		 &		  &		 &		  &		 &  RRab  \\
V69  &  56.155296  & $-$43.549762  &  0.5391860  & 23.23 & 0.79 & 23.20 & 1.01 & 23.31 & 1.49 &		  &		 &		  &		 &		  &		 &  RRab  \\
\hline
\end{tabular}
}
\begin{tablenotes}
\item $\Delta$ refers to the amplitude in order to avoid confusion with the extinction coefficients.
\item RRab* refers to RRab stars that seem to be affected by Blazhko effect.
\item The \textit{peculiar} RRL stars are brighter than the bulk of RRL stars  \citep[see][]{MartinezVazquez2016b}.
\end{tablenotes}
\end{table*}

%% file: comments_vartab.tex
\begin{table*}
\caption{Comments on individual variable stars in Eri~II}
\label{tab:comments}
\begin{tabular}{|l|l|l|}
\hline
\multicolumn{1}{|c|}{Name} &
\multicolumn{1}{|c|}{ID$_{S20}^{(a)}$} &
\multicolumn{1}{l|}{Comment} \\
\hline
 V03 &	& RRd? nosiy	\\
 V04 & 1410944738 &  \\	
 V05 &	& few outliers at minimum light in $i$, bad photometry \\	
 V09 &	& RRd? noisy	\\
 V10 &	& noisy, especially in $i$ \\
 V11 &	& noisy in $i$	\\
 V12 &	& noisy at minimum light in $r$ and $i$\\
 V18 &  & $F475W$ data only samples the minimum\\
 V20 &	& RRd, noisy specially in $i$\\	
 V21 &	& RRd? \\
 V22 &	& Blazhko? $F475W$ data only samples the minimum \\
 V23 &  & detected in ACS, blended with a brighter star in Goodman and DECam \\
 V24 &	& RRd	\\
 V26 &  1410941177  &	\\	
 V27 &	& noisy in $i$\\	
 V30 &	& Blazhko? $F475W$ data only samples the minimum \\
 V31 & 1410943700 &	\\	
 V32 &	& $F475W$ data only samples the minimum \\
 V34  & 1410944785  &	\\	
 V37 & 	& Blazhko	\\
 V38 &	& noisy in $i$ (at maximum light)\\	
 V39 &	& Blazhko?\\	
 V40 &	& Blazhko?\\	
 V45 &	& outliers in $r$, noisy in $i$	\\
 V46 &	& noisy in $i$ and $r$, blend? \\
 V49 & 1410942171 &	\\	
 V50 & 1410940364 & \\	
 V52 &	& peculiar RRL, brighter than the bulk of RRL stars \\	
 V53 &	& Blazhko? few outliers in $r$ and $i$	\\
 V54 &	& noisy in $r$ and $i$	\\
 V56 &	& noisy in $r$, $i$, $F606$ and $F814W$. $F475W$ data only samples the minimum.\\
 V57 &	& RRd? noisy. Uncertain fundamental mode period. The period reported seems to correspond to the first overtone pulsation.\\
 V58 &	& RRd? noisy \\
 V62 &	& RRd? \\
 V63 &	1410937190 & \\	
\hline
\end{tabular}

\begin{tablenotes}
\item $^{(a)}$ Identifier (\verb+'COADD_OBJECT_ID'+) in \citet{Stringer2020} for the seven variable stars of Eridanus~II detected in their work.
\end{tablenotes}

\end{table*}

%% file: photometry_variables_EriII_tab_red.tex
\begin{table*}
\begin{scriptsize}
\centering
\caption{Photometry of the variable stars in Eri~II.}
\label{tab:photometry}
\resizebox{1.0\textwidth}{!}{ 
\begin{tabular}{cccccccccccccccccc} 
\hline
HJD$_i^*$    &    $i$    &    $\sigma_i$    &    HJD$_r^*$    &    $r$    &    $\sigma_r$    &    HJD$_g^*$    &    $g$    &    $\sigma_g$    &  HJD$_{F814W}^*$    &    $F814W$    &    $\sigma_{F814W}$    &    HJD$_{F606W}^*$    &    $F606W$    &    $\sigma_{F606W}$    &    HJD$_{F475W}^*$    &    $F475W$    &    $\sigma_{F475W}$  \\
\hline
\multicolumn{18}{c}{\textbf{V13}} \\
\hline
   58533.1017  &   23.427  &    0.136  &   58537.1070  &   22.573  &    0.046  &   58453.2441  &   23.525  &    0.050  &   57416.3884  &   23.015  &    0.025  &   57416.5219  &   23.532  &    0.023  &   57654.7469  &   23.827  &    0.014 \\
   58503.0564  &   23.510  &    0.101  &   58529.1221  &   22.979  &    0.067  &   58424.2289  &   24.008  &    0.071  &   57416.4046  &   23.075  &    0.020  &   57416.5382  &   23.531  &    0.014  &   57654.8263  &   23.587  &    0.051 \\
   58446.0270  &   22.910  &    0.085  &   58537.0506  &   23.589  &    0.061  &   58424.2475  &   23.872  &    0.072  &   57416.4531  &   22.923  &    0.085  &   57416.5928  &   23.601  &    0.020  &   57654.8807  &   23.062  &    0.054 \\
   58446.1050  &   23.016  &    0.126  &   58529.0551  &   22.539  &    0.036  &   58424.2656  &   24.008  &    0.055  &   57416.4704  &   23.116  &    0.027  &   57416.6626  &   22.479  &    0.075  &   57653.6159  &   23.833  &    0.037 \\
   58446.2993  &   23.271  &    0.103  &   58424.2693  &   23.705  &    0.116  &   58424.2887  &   24.103  &    0.088  &   57422.3494  &   23.095  &    0.025  &   57417.4481  &   23.480  &    0.027  &   57653.6995  &   23.841  &    0.043 \\
   58501.0638  &   23.098  &    0.099  &   58423.1341  &   23.518  &    0.049  &   58424.3077  &   23.887  &    0.075  &   57422.3656  &   23.135  &    0.023  &   57417.4644  &   23.508  &    0.013  &   57653.7484  &   23.610  &    0.110 \\
   58501.1951  &   23.541  &    0.164  &   58423.1678  &   23.509  &    0.053  &   58424.3283  &   24.004  &    0.120  &   57422.4141  &   23.151  &    0.026  &   57417.5136  &   23.606  &    0.017  &   57653.7418  &   23.597  &    0.196 \\
   58502.0657  &   23.122  &    0.062  &   58423.1957  &   23.552  &    0.054  &   58453.0798  &   22.908  &    0.031  &   57422.4314  &   23.124  &    0.027  &   57417.5310  &   23.639  &    0.017  &               &           &          \\
   58537.0941  &   23.100  &    0.068  &   58423.2225  &   23.643  &    0.047  &   58453.0907  &   22.968  &    0.023  &   57423.9390  &   23.057  &    0.027  &   57403.2067  &   22.720  &    0.022  &               &           &          \\
   58503.0434  &   23.372  &    0.105  &   58423.2492  &   23.582  &    0.042  &   58453.1195  &   23.035  &    0.029  &   57423.9552  &   23.080  &    0.016  &   57403.2704  &   23.082  &    0.022  &               &           &          \\
   58504.0459  &   23.532  &    0.215  &   58423.2799  &   23.439  &    0.108  &   58453.1444  &   23.082  &    0.031  &   57424.0037  &   23.160  &    0.028  &               &           &           &               &           &          \\
   58505.0411  &   23.451  &    0.126  &   58423.3190  &   23.504  &    0.076  &   58453.1964  &   23.417  &    0.040  &   57424.0210  &   23.156  &    0.024  &               &           &           &               &           &          \\
   58529.0422  &   22.606  &    0.048  &   58423.3500  &   23.529  &    0.056  &   58424.1685  &   23.760  &    0.064  &   57418.4417  &   22.865  &    0.031  &               &           &           &               &           &          \\
   58529.1091  &   23.041  &    0.068  &   58424.0508  &   23.162  &    0.044  &   58453.2202  &   23.520  &    0.042  &   57418.4579  &   22.866  &    0.035  &               &           &           &               &           &          \\
   58537.0377  &   23.473  &    0.088  &   58424.0832  &   23.251  &    0.046  &   58424.2109  &   23.822  &    0.066  &   57418.5081  &   22.968  &    0.022  &               &           &           &               &           &          \\
   58502.1928  &   23.073  &    0.095  &   58424.1112  &   23.220  &    0.048  &   58423.0999  &   23.812  &    0.056  &   57418.5795  &   23.112  &    0.021  &               &           &           &               &           &          \\
   58423.0271  &           &           &   58424.1482  &   23.463  &    0.040  &   58453.3123  &   23.673  &    0.062  &   57654.7640  &   23.052  &    0.041  &               &           &           &               &           &          \\
   58423.2397  &   23.455  &    0.065  &   58424.1755  &   23.497  &    0.045  &   58453.2678  &   23.604  &    0.053  &   57654.8092  &   23.030  &    0.017  &               &           &           &               &           &          \\
   58423.2658  &           &           &   58424.1956  &   23.569  &    0.054  &   58424.1323  &   23.674  &    0.044  &   57653.6330  &   23.002  &    0.024  &               &           &           &               &           &          \\
   58424.1839  &   23.463  &    0.078  &   58424.2164  &   23.568  &    0.063  &   58424.1033  &   23.462  &    0.054  &   57653.6824  &   23.052  &    0.023  &               &           &           &               &           &          \\
   58423.3364  &   23.605  &    0.095  &   58453.3514  &   23.609  &    0.102  &   58424.0746  &   23.373  &    0.032  &   57653.7655  &   22.881  &    0.027  &               &           &           &               &           &          \\
   58424.0622  &   23.174  &    0.083  &   58453.3210  &   23.544  &    0.055  &   58424.0405  &   23.177  &    0.037  &   57654.7998  &   23.127  &    0.198  &               &           &           &               &           &          \\
   58424.0943  &   23.297  &    0.085  &   58453.2753  &   23.386  &    0.044  &   58423.3281  &   23.914  &    0.067  &   57653.6730  &   23.111  &    0.271  &               &           &           &               &           &          \\
   58424.1213  &   23.122  &    0.065  &   58424.2352  &   23.732  &    0.056  &   58423.2922  &   23.793  &    0.077  &               &           &           &               &           &           &               &           &          \\
   58423.1552  &   23.522  &    0.120  &   58453.2264  &   23.374  &    0.043  &   58423.2574  &   23.784  &    0.041  &               &           &           &               &           &           &               &           &          \\
   58423.1207  &   23.423  &    0.080  &   58453.2029  &   23.176  &    0.035  &   58423.2307  &   23.831  &    0.041  &               &           &           &               &           &           &               &           &          \\
   58423.0669  &           &           &   58453.1518  &   23.123  &    0.035  &   58423.2031  &   23.878  &    0.039  &               &           &           &               &           &           &               &           &          \\
   58424.1595  &   23.285  &    0.056  &   58453.1270  &   23.021  &    0.025  &   58423.1758  &   23.852  &    0.046  &               &           &           &               &           &           &               &           &          \\
   58423.3067  &   23.291  &    0.076  &   58453.0999  &   22.979  &    0.027  &   58423.1456  &   23.864  &    0.066  &               &           &           &               &           &           &               &           &          \\
   58453.1365  &   23.044  &    0.057  &   58424.3132  &   23.519  &    0.072  &   58453.3445  &   24.135  &    0.156  &               &           &           &               &           &           &               &           &          \\
   58453.1841  &   23.179  &    0.048  &   58423.0865  &   23.417  &    0.062  &   58424.1901  &   23.841  &    0.049  &               &           &           &               &           &           &               &           &          \\
   58424.2037  &   23.391  &    0.073  &   58424.2940  &   23.622  &    0.081  &               &           &           &               &           &           &               &           &           &               &           &          \\
   58424.2232  &   23.454  &    0.105  &   58424.2530  &   23.590  &    0.067  &               &           &           &               &           &           &               &           &           &               &           &          \\
   58424.2418  &   23.565  &    0.097  &   58453.2499  &   23.262  &    0.062  &               &           &           &               &           &           &               &           &           &               &           &          \\
   58424.2596  &   23.577  &    0.091  &               &           &           &               &           &           &               &           &           &               &           &           &               &           &          \\
   58424.2823  &   23.388  &    0.069  &               &           &           &               &           &           &               &           &           &               &           &           &               &           &          \\
   58424.3007  &   23.678  &    0.106  &               &           &           &               &           &           &               &           &           &               &           &           &               &           &          \\
   58424.3215  &   23.671  &    0.129  &               &           &           &               &           &           &               &           &           &               &           &           &               &           &          \\
   58453.3367  &   23.427  &    0.053  &               &           &           &               &           &           &               &           &           &               &           &           &               &           &          \\
   58453.2851  &   23.388  &    0.076  &               &           &           &               &           &           &               &           &           &               &           &           &               &           &          \\
   58453.2583  &   23.352  &    0.057  &               &           &           &               &           &           &               &           &           &               &           &           &               &           &          \\
   58453.2358  &   23.462  &    0.063  &               &           &           &               &           &           &               &           &           &               &           &           &               &           &          \\
   58453.1097  &   23.087  &    0.060  &               &           &           &               &           &           &               &           &           &               &           &           &               &           &          \\
   58453.2115  &   23.240  &    0.055  &               &           &           &               &           &           &               &           &           &               &           &           &               &           &          \\
   58423.2130  &   23.583  &    0.076  &               &           &           &               &           &           &               &           &           &               &           &           &               &           &          \\
   58423.1864  &   23.624  &    0.090  &               &           &           &               &           &           &               &           &           &               &           &           &               &           &          \\
\hline
\hline
\end{tabular}
}
\begin{tablenotes}
\item $^*$Heliocentric Julian Date of mid-exposure minus 2,400,000 days.
\item This table is a portion of its entirely form which will be available as Supplementary material (online) in the journal.
\end{tablenotes}
\end{scriptsize}
\end{table*}